\documentclass{emulateapj}

\usepackage{natbib,aas_macros}
\usepackage{multirow,color}

\usepackage{comment}

\usepackage{color}

\slugcomment{ApJ in Press}

\shorttitle{Subaru/HSC Ly$\alpha$ Luminosity Function at $z=7.0$}
\shortauthors{Itoh et al.}

\begin{document}

\title{CHORUS II. Subaru/HSC Determination of the Ly$\alpha$ Luminosity Function 
at $z=7.0$: Constraints on Cosmic Reionization Model Parameter\\
}

\author{
Ryohei~Itoh\altaffilmark{1,2}, 
Masami~Ouchi\altaffilmark{1,3}, 
Haibin~Zhang\altaffilmark{1,2}, 
Akio~K.~Inoue\altaffilmark{4}, 
Ken~Mawatari\altaffilmark{4}, 
Takatoshi~Shibuya\altaffilmark{13}, 
Yuichi~Harikane\altaffilmark{1,2}, 
Yoshiaki~Ono\altaffilmark{1}, 
Haruka~Kusakabe\altaffilmark{5}, 
Kazuhiro~Shimasaku\altaffilmark{5, 6}, 
Seiji~Fujimoto\altaffilmark{1},
Ikuru~Iwata\altaffilmark{8, 9}, 
Masaru~Kajisawa\altaffilmark{10}, 
Nobunari~Kashikawa\altaffilmark{7, 9}, 
Satoshi~Kawanomoto\altaffilmark{7}, 
Yutaka Komiyama\altaffilmark{7, 11}, 
Chien-Hsiu~Lee\altaffilmark{8}, 
Tohru~Nagao\altaffilmark{10}, 
and Yoshiaki~Taniguchi\altaffilmark{12} 
}
\email{itoh@icrr.u-tokyo.ac.jp}
\altaffiltext{1}{Institute for Cosmic Ray Research, The University of Tokyo, 5-1-5 Kashiwanoha, Kashiwa,
Chiba 277-8582, Japan}
\altaffiltext{2}{Department of Physics, Graduate School of Science, The University of Tokyo, 7-3-1 Hongo, Bunkyo-ku, Tokyo 113-0033, Japan}
\altaffiltext{3}{Kavli Institute for the Physics and Mathematics of the Universe (Kavli IPMU, WPI), The University of Tokyo, 5-1-5 Kashiwanoha, Kashiwa, Chiba 277-8583, Japan}
\altaffiltext{4}{Department of Environmental Science and Technology, Faculty of Design Technology, Osaka Sangyo University, 3-1-1 Nagaito, Daito, Osaka 574-8530, Japan}
\altaffiltext{5}{Department of Astronomy, Graduate School of Science, The University of Tokyo, 7-3-1 Hongo, Bunkyo-ku, Tokyo 113-0033, Japan}
\altaffiltext{6}{Research Center for the Early Universe, Graduate School of Science, The University of Tokyo, 7-3-1 Hongo, Bunkyo, Tokyo 113-0033, Japan}
\altaffiltext{7}{National Astronomical Observatory of Japan, 2-21-1 Osawa, Mitaka, Tokyo 181-8588, Japan}
\altaffiltext{8}{Subaru Telescope, National Astronomical Observatory of Japan, 650 N Aohoku Pl, Hilo, HI 96720}
\altaffiltext{9}{7Department of Astronomy, School of Science, Graduate University for Advanced Studies (SOKENDAI), 2-21-1, Osawa, Mitaka, Tokyo 181-8588, Japan}
\altaffiltext{10}{Research Center for Space and Cosmic Evolution, Ehime University, 2-5 Bunkyo-cho, Matsuyama, Ehime 790-8577, Japan}
\altaffiltext{11}{Graduate University for Advanced Studies (SOKENDAI), 2-21-1 Osawa, Mitaka, Tokyo 181-8588, Japan}
\altaffiltext{12}{The Open University of Japan, Wakaba 2-11, Mihama-ku, Chiba 261-8586, Japan}
\altaffiltext{13}{Kitami Institute of Technology, 165, Koen-cho, Kitami, Hokkaido 090-8507, Japan}

\begin{abstract}
We present the Ly$\alpha$ luminosity function (LF) derived from 
34 Ly$\alpha$ emitters (LAEs) at $z=7.0$ 
on the sky of $3.1$ deg$^2$, the largest
sample compared to those in the literature obtained at a redshift $z\gtrsim7$. The LAE sample is made by 
deep large-area Subaru narrowband observations conducted
by the Cosmic HydrOgen Reionization Unveiled with Subaru (CHORUS) project.
The $z=7.0$ Ly$\alpha$ LF of our project is consistent with those of the previous
DECam and Subaru studies at the bright and faint ends, respectively,
while our $z=7.0$ Ly$\alpha$ LF has uncertainties significantly smaller
than those of the previous study results. Exploiting the small errors of our measurements,
we investigate the shape of the faint to bright-end Ly$\alpha$ LF. We find
that the $z=7.0$ Ly$\alpha$ LF shape can be explained by
the steep slope of $\alpha \simeq -2.5$ suggested at $z=6.6$,
and that there is no clear signature of a bright-end excess at $z\simeq 7$ 
claimed by the previous work, which was thought to be made by the ionized bubbles
around bright LAEs whose Ly$\alpha$ photons could easily escape from
the partly neutral IGM at $z \simeq 7$. We estimate the Ly$\alpha$ luminosity
densities (LDs) with Ly$\alpha$ LFs at $z\simeq 6-8$ 
given by our and the previous studies, and compare the evolution of the UV-continuum LD estimated with dropouts.
The Ly$\alpha$ LD monotonically decreases from $z\sim 6$ to $8$,
and evolves stronger than the UV-continuum LD, indicative of the Ly$\alpha$ damping wing absorption
of the IGM towards the heart of the reionization epoch.
\end{abstract}


\section{Introduction}
Cosmic reionization is one of the most important events
in the early history of the universe, 
as massive stars and/or active galactic nuclei ionize 
the neutral hydrogen in the intergalactic medium (IGM).
It is suggested that the cosmic reionization has been completed by $z\sim6$
from the observations of the Gunn-Peterson trough in quasar spectra \citep{fan2006, goto2011}
and analysis of gamma-ray burst damping wing absorptions 
\citep{totani2006, totani2014, totani2016, chornock2013, mcgreer2015}. 
The Thomson scattering optical depth of the cosmic microwave background 
indicates that the cosmic reionization event takes place at $7<z<10$ \citep{planck2016a}. 

Ly$\alpha$ emitters (LAEs) are used as a tool for probing the cosmic reionization. 
The LAE population can be characterized
by the Ly$\alpha$ luminosity function (LF). 
The Ly$\alpha$ LFs are often fit with a Schechter function parametrized 
by the characteristic number density $\phi^*$, 
the characteristic luminosity $L^*$, 
and the faint-end slope $\alpha$ \citep{schechter1976}.
Three Schechter parameters are used 
to investigate the redshift evolution of the Ly$\alpha$ LF. 

Previous narrowband ($NB$) studies reveal that Ly$\alpha$ LFs 
do not evolve from $z=3$ to $z=5.7$ \citep{ouchi2008},
and decrease from $z=5.7$ to $z=6.6$ 
\citep{kashikawa2006, hu2010, ouchi2010, kashikawa2011, santos2016, konno2018}.
The decrease of Ly$\alpha$ LFs at $z=5.7-6.6$ 
is too large to be explained by the decrease of 
UV LFs estimated with dropouts, 
which correlates with 
the star formation rate density. 
Because Ly$\alpha$ photons are resonantly scattered by 
neutral hydrogen in the IGM, 
it is suggested that the increase of the Ly$\alpha$ damping wing absorption 
of the IGM is needed to explain the decrease of the Ly$\alpha$ LFs. 
\cite{konno2014} investigate the Ly$\alpha$ LF at $z=7.3$, 
and identify that the Ly$\alpha$ LF declines from $z=6.6$ to $7.3$ 
more rapidly than from $z=5.7$ to $z=6.6$, 
possibly due to the accelerated increase of the neutral hydrogen fraction 
at a given redshift interval.

\cite{konno2018} derive the Ly$\alpha$ LFs using the largest $z=5.7$ and $6.6$ LAE samples, to date, 
obtained by SILVERRUSH program (\citealt{ouchi2018})
with the Subaru/Hyper Suprime-Cam 
(HSC; \citealt{miyazaki2018, komiyama2018, kawanomoto2017, furusawa2018}) 
survey data. 
The total areas of the HSC survey are 
$13.8\ {\rm deg}^2$ and $21.2\ {\rm deg}^2$
for $z=5.7$ and 6.6 LAEs, respectively .
Exploiting the large area of the sky coverage, 
the HSC survey reaches the bright luminosity limit of $\log L_{{\rm Ly\alpha}}[{\rm erg\ s^{-1}}]=43.8$. 
\cite{konno2018} use the LAE samples 
of the HSC survey and the previous observations \citep{ouchi2008,ouchi2010} 
to derive the best-fit Schechter parameters. 
\cite{konno2018} obtain the best-fit values of $\alpha=-2.6$ and $-2.5$ for 
the Ly$\alpha$ LFs at $z=5.7$ and $z=6.6$, respectively, 
which are steeper than those of the UV LFs at these redshifts \cite[e.g.,][]{bouwens2015b}.
Similar $\alpha$ values for the Ly$\alpha$ LF are also 
given by the spectroscopic search reaching luminosities fainter 
than $L^*$ (\citealt{drake2017b};
see also 
\citealt{rauch2008, martin2008, cassata2011, henry2012, dressler2011, dressler2015}). 
\cite{konno2018} also argue that 
the bright-end of the LFs may have some systematic effects 
such as the contribution from AGNs, blended merging galaxies, and/or large ionized bubbles around the bright LAEs 
\citep[see also][]{matthee2015, santos2016}. 

Ly$\alpha$ LFs at $z\simeq7.0$ are investigated by \cite{zhengy2017} and \cite{ota2017}. 
\cite{zhengy2017} use an $NB$ filter, 
$NB964$ ($\lambda_c = 9642\ {\rm \AA}$, ${\rm FWHM}=90\ {\rm \AA}$), 
installed on the Dark Energy Camera (DECam) on the NOAO/CTIO 4 m Blanco telescope. 
\cite{zhengy2017} identify 23 LAE candidates at $z=6.9$ in a $2\ {\rm deg^2}$ sky 
of the  Cosmic Evolution Survey (COSMOS) field.
The Ly$\alpha$ LF at $z=6.9$ is comparable to the one at $z=7.3$ \citep{konno2014} 
at the relatively faint end, $\log L_{{\rm Ly\alpha}} [{\rm erg\ s^{-1} }]<43.0$, 
showing a significant drop from the one at $z=6.6$ \cite{konno2018}.
The Ly$\alpha$ LF of \cite{zhengy2017} shows 
a significant bright-end excess over the best-fit Schechter function,
which cannot be explained by the shape of the Schechter function. 
\cite{zhengy2017} discuss that the bright-end excess 
is an indicator of large ionized bubbles around bright LAEs 
during the epoch of reionization \citep[EoR; e.g.,][]{santos2016, bagley2017, konno2018}.
\cite{ota2017} detect 20 LAEs at $z=7.0$ in the total area of $0.5\ {\rm deg^2}$ 
in the Subaru/XMM-Newton Deep Survey (SXDS) and Subaru Deep Field (SDF) fields
using Subaru Telescope Suprime-Cam $NB973$ 
($\lambda_c = 9755\ {\rm \AA}$, ${\rm FWHM}=200\ {\rm \AA}$; hereafter $NB973_{\rm SC}$).
\cite{ota2017} find that the Ly$\alpha$ LF evolves 
moderately from $z=6.6$ to $7.0$ and more rapidly from $z=7.0$ to $7.3$. 
\cite{ota2017} compare the observed Ly$\alpha$ LF 
with the one predicted from the LAE evolution model, 
and claim that the neutral hydrogen fraction increases rapidly at $z>6$. 

There are two discrepancies of the Ly$\alpha$ LFs at $z\simeq7$ 
between \cite{zhengy2017} and \cite{ota2017}. 
At the bright end $\log L_{{\rm Ly\alpha}}[{\rm erg\ s^{-1}}] > 43.2$, 
the data points of \cite{ota2017} fall below those of \cite{zhengy2017}. 
On the other hand, at the faint end $\log L_{{\rm Ly\alpha}}[{\rm erg\ s^{-1}}] < 43.2$, 
the data points of \cite{ota2017} exceed those of \cite{zhengy2017}. 
The other discrepancy is the existence of the bright-end excess. 
The Ly$\alpha$ LF of \cite{zhengy2017} shows a clear bright-end excess 
over the best-fit Schechter function, 
while that of \cite{ota2017} does not have such a significant excess.  

The origin of these discrepancies are unclear. 
The possible explanation of the bright-end LF discrepancy is that 
the survey volume of \cite{ota2017} may not be enough 
to identify the bright-end excess of the Ly$\alpha$ LF. 
\cite{ota2017} cover the sky of $0.5$ deg$^2$, 
that is 4 times smaller than that of \cite{zhengy2017}. 
The potential reason of the faint-end LF discrepancy is 
that the data of \cite{zhengy2017} may not be 
deep enough to determine the faint end of the Ly$\alpha$ LF. 
The exposure time of \cite{zhengy2017} is 34 hours with 4 m Blanco telescope, 
while \cite{ota2017} reach the exposure time of 60 hours with 8 m Subaru/Suprime-Cam. 
Thus, deeper and larger-area LAE surveys are needed 
to resolve these discrepancies. 

This paper is one in a series of papers from the program named 
Cosmic HydrOgen Reionization Unveiled with Subaru (CHORUS; PI: A. K. Inoue).
CHORUS is the series of deep HSC imaging observations 
with five custom narrowband filters: 
$NB387$, $NB527$, $NB718$, $IB945$, and $NB973$, 
which are not included in the HSC Subaru Strategic Program (SSP) survey data. 
CHORUS provides the legacy data of large-area and deep 
$NB$ images that allow us to make statistical samples of LAEs 
at $z= 3.3,\ 4.9,\ 6.8,$ and $7.0$. 
In this paper, we present the results of the $z=7.0$ LAEs. 
In the survey volumes mostly independent of 
\cite{zhengy2017} and \cite{ota2017}, 
we derive the bright-end of the $z=7.0$ Ly$\alpha$ LF to test the existence of the bright-end excess. 
We also study the faint end of the Ly$\alpha$ LF 
that remains the problem, 
the discrepancy between \cite{ota2017} and \cite{zhengy2017}.
In section \ref{sec:observations}, we describe the details of our $z=7.0$ LAE survey 
and the selection of our LAE candidates.
In section \ref{sec:luminosity_function}, 
we derive the Ly$\alpha$ LF at $z=7.0$, 
and compare with those obtained by previous studies. 
In section \ref{sec:discussion}, 
we discuss the evolution of the Ly$\alpha$ LFs at $z\sim7$ 
and cosmic reionization. 
Throughout this paper, we adopt AB magnitudes \citep{oke1974} 
and a concordance cosmology with $(\Omega_m,\ \Omega_{\Lambda},\ h,\ \sigma_8)=(0.3,\ 0.7,\ 0.7,\ 0.8)$ 
consistent with the constraints by the recent $WMAP$ and $Planck$ observations \citep{hinshaw2013, planck2016a}.

\section{Observations and Data Reduction}
\label{sec:observations}

\subsection{CHORUS $NB973$ Imaging}
\label{sec:imaging_observations}
\begin{figure}
\epsscale{1.25}
\plotone{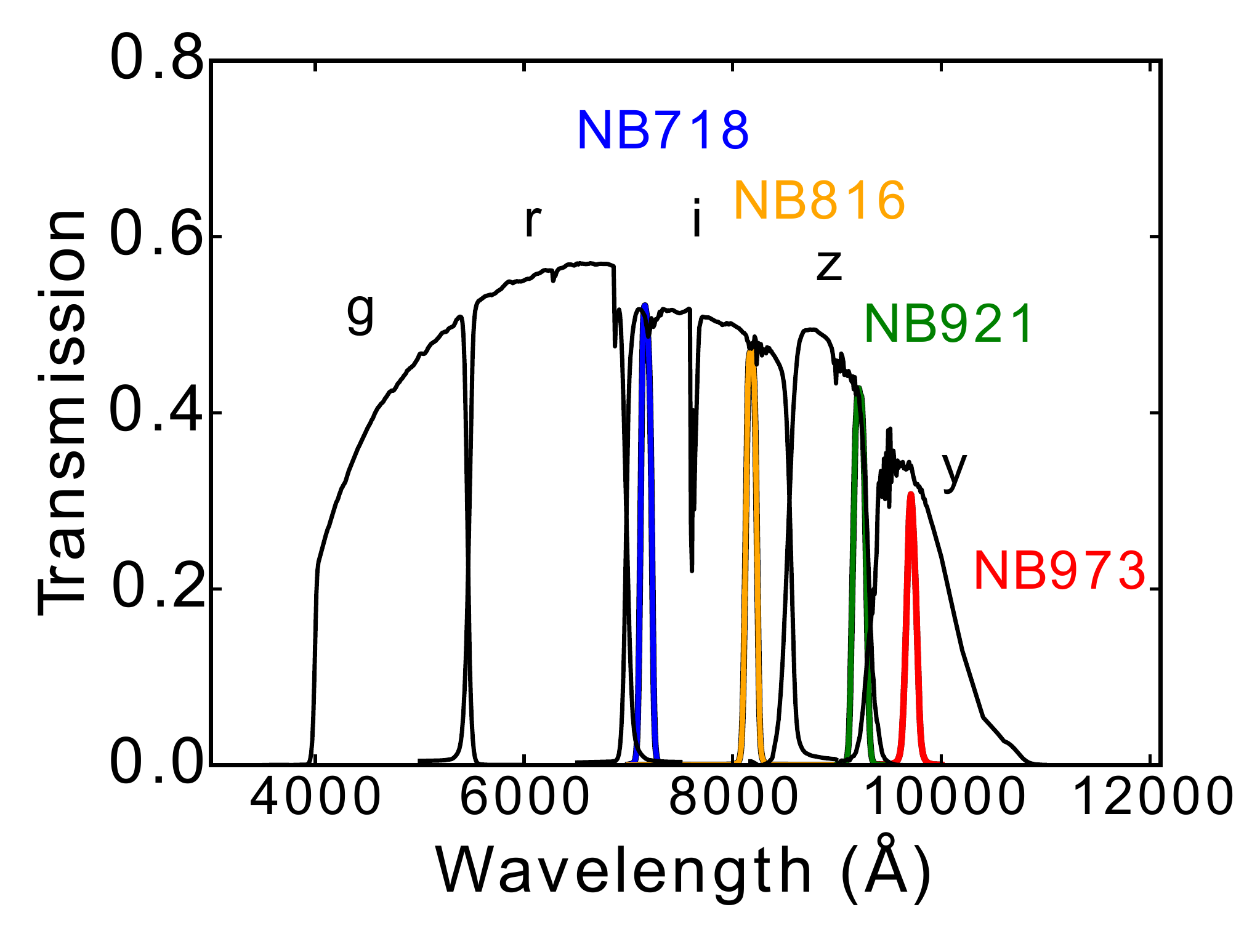}
\caption{
Filter response of $NB973_{\rm HSC}$ is shown with the red line. 
The other colored and black lines represent the response of 
other HSC $NB$ and $BB$ filters respectively. 
These response curves include the CCD quantum efficiency of HSC, 
airmass, the transmittance of the dewar window and 
primary focus unit, and the reflectivity of the primary mirror.
\label{fig:response}}
\end{figure}

The HSC $NB973$ band (hereafter $NB973_{{\rm HSC}}$) has a central wavelength of $\lambda_c = 9715$ \AA\ and an FWHM of 100 \AA\ 
to identify LAEs in the redshift range of $z = 6.95 - 7.03$. 
We show the response curves of $NB973_{{\rm HSC}}$ and the other $NB$ and broadband ($BB$) filters in Figure \ref{fig:response}.
Note that $NB973_{\rm SC}$ has the central wavelength of $\lambda_c=9755\ {\rm \AA}$ 
and an FWHM of $200\ {\rm \AA}$ \citep{ota2017},
which is broader than our $NB973_{\rm HSC}$.
We carried out $NB973_{\rm HSC}$ observations in 2017 January 27 and 29 in two fields, COSMOS and SXDS. 
Table \ref{tab:data_summary} shows the details of our $NB973_{\rm HSC}$  imaging data and other band data used in this study.

In our $NB973_{\rm HSC}$ images, we mask out regions contaminated with diffraction spikes and halos of bright stars using bright star masks provided by \cite{coupon2018}.
We do not use regions affected by sky over- and under-subtractions around large objects.
After the removal of these regions, the effective survey areas (volumes) of $NB973_{\rm HSC}$ images are 1.64 ${\rm deg}^2$ (1.15 $\times 10^6\ {\rm Mpc^3}$) and 1.50 ${\rm deg}^2$ (1.04 $\times 10^6\ {\rm Mpc^3}$) in the COSMOS and SXDS fields, respectively. 
The survey volume of our study has an overlap with 
those of \cite{ota2017} and \cite{zhengy2017}. 
Approximately $20\%$ of our survey volume overlap with that of \cite{zhengy2017}. 
There is the overlap of $\sim8\%$ in the survey volume of our and \cite{ota2017} observations. 
The total survey area (volume) is larger than those of \cite{zhengy2017} and \cite{ota2017}.
The total exposure times are 14.7 hours in the COSMOS field and 4.7 hours in the SXDS field. 

\subsection{Data Reduction}
\label{sec:data_reduction}

Our $NB973_{\rm HSC}$ data are reduced 
with {\tt hscPipe}\footnote[1]{http://hsc.mtk.nao.ac.jp/pipedoc\_e/}  \citep{bosch2018} version 4.0.5, 
which is based on the Large Synoptic Survey Telescope (LSST) pipeline 
\citep{ivezic2008, axelrod2010, juric2015}.
The {\tt hscPipe} performs CCD-by-CCD reduction, calibration 
for astrometry, and photometric zero point determination.
The astrometry and photometric zero point are obtained based on the data 
from the Panoramic Survey Telescope and Rapid Response System 1 imaging survey 
(PanSTARRS1; \citealt{schlafly2013, tonry2012, magnier2013}). 

The photometric zero points and the color-term coefficients $(a,\ b,\ c)$ are defined as
$NB973_{{\rm HSC} }= y_{{\rm PS1}} + a + b \times (y_{{\rm PS1}}-z_{{\rm PS1}}) + c \times (y_{{\rm PS1}}-z_{{\rm PS1}})^2$,
where $z_{{\rm PS1}}$ and $y_{{\rm PS1}}$ are the $z$- and $y$-band magnitudes in a $2''.0$ diameter aperture in PanSTARRS catalog. 
$NB973_{{\rm HSC}}$ is the $NB973_{\rm HSC}$ magnitude in a $2''.0$ diameter aperture in our images. 
Note that the seeing sizes of PanSTARRS1 $z$ and $y$ images are $\approx1''$.
We determine the color-term coefficients using the spectra of 175 Galactic stars given in \cite{gunn1983},
and obtain $(a,\ b,\ c) = (-0.00640165,\ -0.03915281,\ -0.24088565)$.

We estimate limiting magnitudes of our images with the {\tt limitmag} task 
in the Suprime-Cam Deep field REDuction package (SDFRED; \citealt{yagi2002, ouchi2004}).
The final $NB973_{\rm HSC}$ images of COSMOS and SXDS fields reach the 5$\sigma$ limiting magnitudes 
of 24.9 and 24.2, respectively, in a $1''.5$ diameter aperture. 
The seeing sizes of the HSC images are typically better than $0''.8$ arcsec. 
If we assume a simple top-hat selection function for LAEs whose redshift distribution is defined by the FWHM of our $NB973_{\rm HSC}$, 
the survey volumes are $1.15 \times 10^6\ {\rm Mpc^3}$ and $1.04 \times 10^6\ {\rm Mpc^3}$ in COSMOS and SXDS, respectively. 
Estimating the total magnitudes of the sources, we use {\tt cmodel} magnitudes defined in the {\tt hscPipe}. 
The {\tt cmodel} magnitude is a weighted combination of exponential and de Vaucouleurs fits to the light profile of each object.
The total magnitudes and colors are corrected for Galactic extinction
\citep{schlegel1998}.

In addition to our $NB973_{\rm HSC}$ imaging data,
we use CHORUS $NB718$ imaging data (H. Zhang et al. in preparation) and
HSC SSP internal release data of S16A \citep{aihara2018b} consisting of 
broadband ($g,\ r,\ i,\ z$, and $y$) and narrowband ($NB816$ and $NB921$) images. 
Note that the CHORUS $NB718$ and HSC SSP imaging data are reduced 
in the same manner as our $NB973_{\rm HSC}$ imaging data. 
The {\tt hscPipe} performs the detections and flux measurements 
of our sources by the method called the {\tt forced} photometry.
In the {\tt forced} photometry,
we estimate the centroid and shape of an object in a reference band,
and measure fluxes in all of the other bands. 
We apply the {\tt forced} photometry for the detections and flux measurements of our sources.
We name these images and source catalogs ``CHORUS version 1.0''.

\begin{deluxetable*}{ccccccccccc}
\tablecolumns{7}
\tabletypesize{\scriptsize}
\tablecaption{Summary of Our Imaging Observations and Data
\label{tab:data_summary}}
\setlength{\tabcolsep}{0.0in}
\tablewidth{0pt}
\tablehead{
\colhead{Field} & 
\colhead{Band} &
\colhead{Exposure Time} & 
\colhead{PSF Size} &
\colhead{Area} &
\colhead{$m_{{\rm lim}}$} &
\colhead{Date of Observation} \\
\colhead{} & 
\colhead{} &
\colhead{(s)} & 
\colhead{(arcsec)} &
\colhead{(deg$^2$)} &
\colhead{(5$\sigma$ AB mag)} &
\colhead{} 
}
\startdata
COSMOS & $NB973_{\rm HSC}$ & 52,800 & 0.64 & 1.64 & 25.0 & 2017 Jan. 27-29 \\ 
 & $NB718$\tablenotemark{a} & 27,600 & 0.69 & 1.64 & 26.2 & 2017 Mar. 23-25\\ \hline
SXDS & $NB973_{\rm HSC}$ & 16,800 & 0.78 & 1.50 & 24.3 & 2017 Jan. 27-29 \\ \hline
\multicolumn{7}{c}{Archival HSC Data (S16A)}\\ \hline
COSMOS & $g$ & & & & 26.9\tablenotemark{b} & \\
& $r$ & & & & 26.6\tablenotemark{b} & \\
& $i$ & & & & 26.2\tablenotemark{b} & \\
& $z$ & & & & 25.8\tablenotemark{b} & \\
& $y$ & & & & 25.1\tablenotemark{b} & \\
& $NB816$ & & & & 25.7\tablenotemark{b} & \\
& $NB921$ & & & & 25.6\tablenotemark{b} & \\ \hline
SXDS & $g$ & & & & 26.9\tablenotemark{b} & \\
& $r$ & & & & 26.4\tablenotemark{b} & \\
& $i$ & & & & 26.3\tablenotemark{b} & \\
& $z$ & & & & 25.6\tablenotemark{b} & \\
& $y$ & & & & 24.9\tablenotemark{b} & \\
& $NB816$ & & & & 25.5\tablenotemark{b} & \\
& $NB921$ & & & & 25.5\tablenotemark{b} & 
\enddata
\tablenotetext{a}{
Although we use the photometric data of $NB718$, 
the details of $NB718$ are discussed in H. Zhang et al. in preparation. 
}
\tablenotetext{b}{
These values are presented in \cite{konno2018}. 
}
\end{deluxetable*}

\subsection{Photometric Sample of $z=7.0$ LAEs }
\label{sec:photometricsample}
\begin{figure}
\begin{minipage}{1\hsize}
\epsscale{1.25}
\plotone{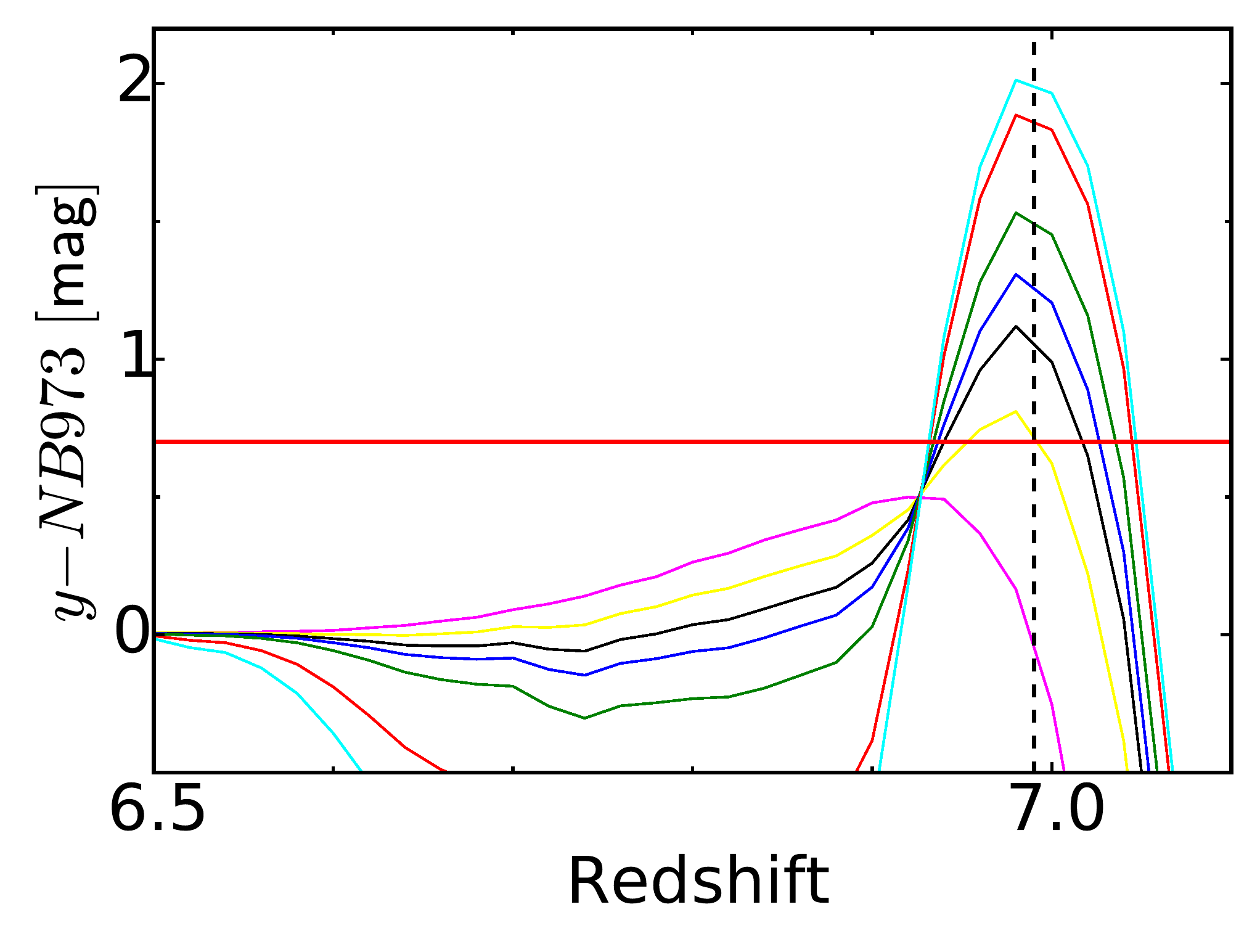}
\end{minipage}
\begin{minipage}{1\hsize}
\epsscale{1.25}
\plotone{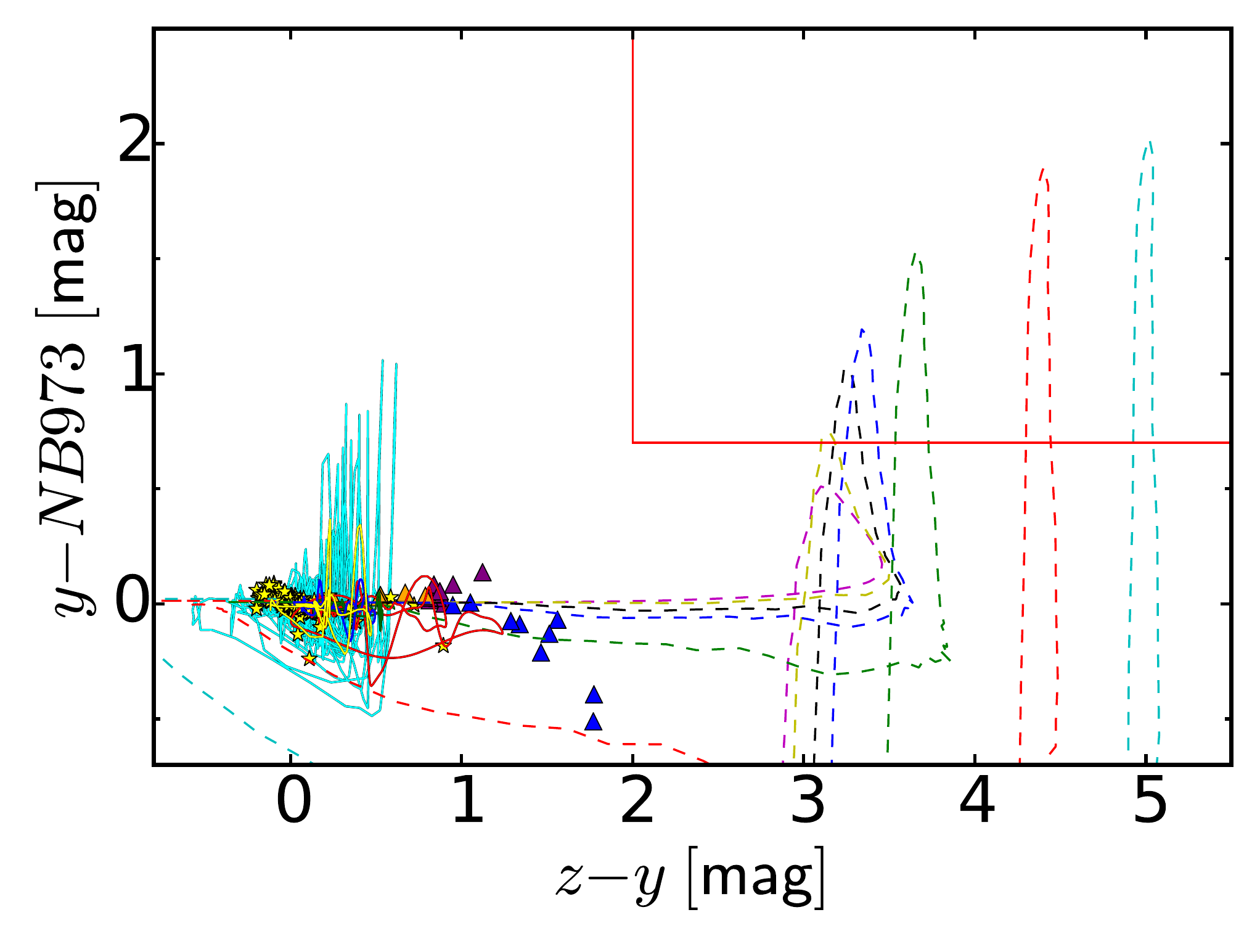}
\caption{
Top: Expected colors of LAEs as a function of redshift.
We show the results for 
$EW_0=0\ {\rm \AA}$ (magenta), $10\ {\rm \AA}$ (yellow), $20\ {\rm \AA}$ (black), $30\ {\rm \AA}$ (blue), $50\ {\rm \AA}$ (green), $150\ {\rm \AA}$ (red), and $300\ {\rm \AA}$ (cyan) cases.
The black dashed line is the redshift of the Ly$\alpha$ emission our $NB973_{\rm HSC}$ band identifies.
The red horizontal line indicates the selection criteria we adopt, $y-NB973_{\rm HSC}>0.7$.
Bottom: Color-color diagram of model LAEs (dashed lines).
The red solid vertical and horizontal lines 
represent the selection criterion we adopt, $y-NB973_{\rm HSC}>0.7$ and $z - y >2.0$.
For comparison, 
we plot the colors of local starburst galaxies using the template spectra of \cite{kinney1996}
with the solid cyan lines.
We also show the colors of E, Sbc, Scd, and Im galaxies using the template spectra of \cite{coleman1980}
with the solid red, green, blue, and yellow lines.
The purple, blue and orange triangles represent the colors of the L/T/M type dwarfs 
calculated from the spectra of 
\cite{burgasser2004, burgasser2006b, burgasser2006a, burgasser2008, burgasser2010} and \cite{kirkpatrick2010} 
from the SpeX Prism Spectral Libraries.
\label{fig:criteria}}
\end{minipage}
\end{figure}

We construct the sample of LAEs at $z=7.0$ based on the narrowband color excess by the Ly$\alpha$ emission, $y-NB973_{\rm HSC}$, 
and no detection of bluer bands. 
To determine the selection criteria for $z=7.0$ LAEs, we predict the expected colors of LAEs. 
We assume a simple model SED of LAEs with a flat continuum ($f_{\nu}={\rm const}$) and a $\delta$-function Ly$\alpha$ emission 
with rest-frame equivalent widths of $EW_0 = {\rm  0,\ 10,\ 20,\ 30,\ 50,\ 150,\ and\ 300\  \AA}$. 
We adopt the UV continuum slope of $\beta = -2$, although $\beta = 0, -1,\ {\rm and}\ -3$ give the similar results.
We redshift the spectra, and apply the IGM absorption described in \citet{madau1995}.
We calculate colors of these LAEs with the response curves of HSC shown in Figure \ref{fig:response}.

The top panel of Figure \ref{fig:criteria} shows the calculated color excess as a function of redshift.
As seen in this color-redshift diagram, $z=7$ LAEs are expected to 
show a narrowband excess of $y - NB973_{\rm HSC}>0.7$, 
if the condition of  $EW_0\gtrsim10\ {\rm \AA}$ is met.
We adopt $y-NB973_{\rm HSC}>0.7$ color as one of our $z=7.0$ LAE selection criteria.
Note that both \cite{ota2017} and \cite{zhengy2017} adopt the narrowband excess of LAEs  
corresponding to $EW_{{\rm Ly\alpha}}\gtrsim10\ {\rm \AA}$, which is similar to ours.
The bottom panel of Figure \ref{fig:criteria} shows color-color diagram of our model LAEs with various $EW_0$s.
We also plot the model colors of potential low redshift interlopers.
As seen in the color-color diagram, our model LAEs exhibit a red $z-y$ color due to the existence of the GP trough.
To remove potential low redshift interlopers, we adopt $z - y>2.0$.

\begin{figure}
\begin{minipage}{1\hsize}
\epsscale{1.25}
\plotone{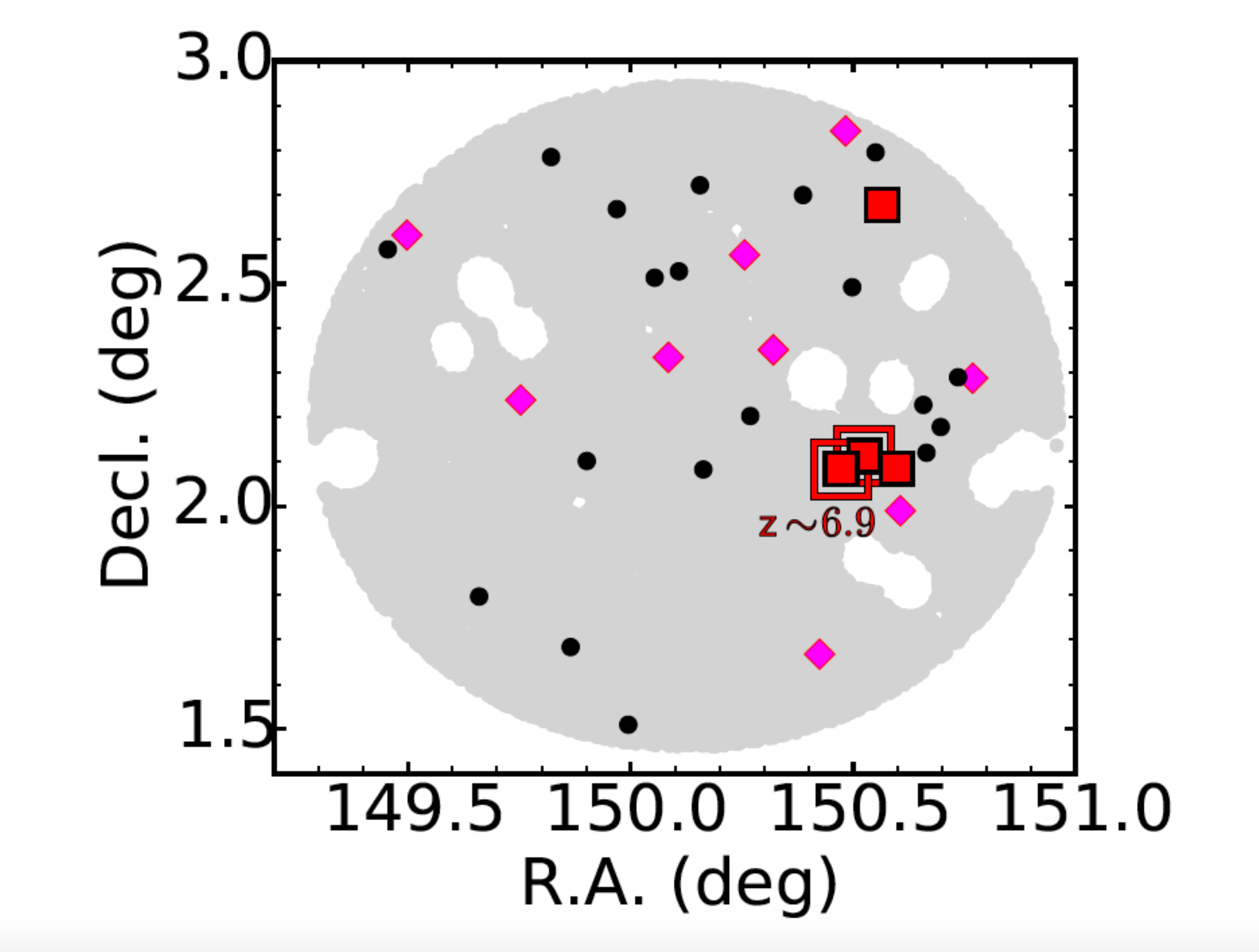}
\end{minipage}
\begin{minipage}{1\hsize}
\epsscale{1.25}
\plotone{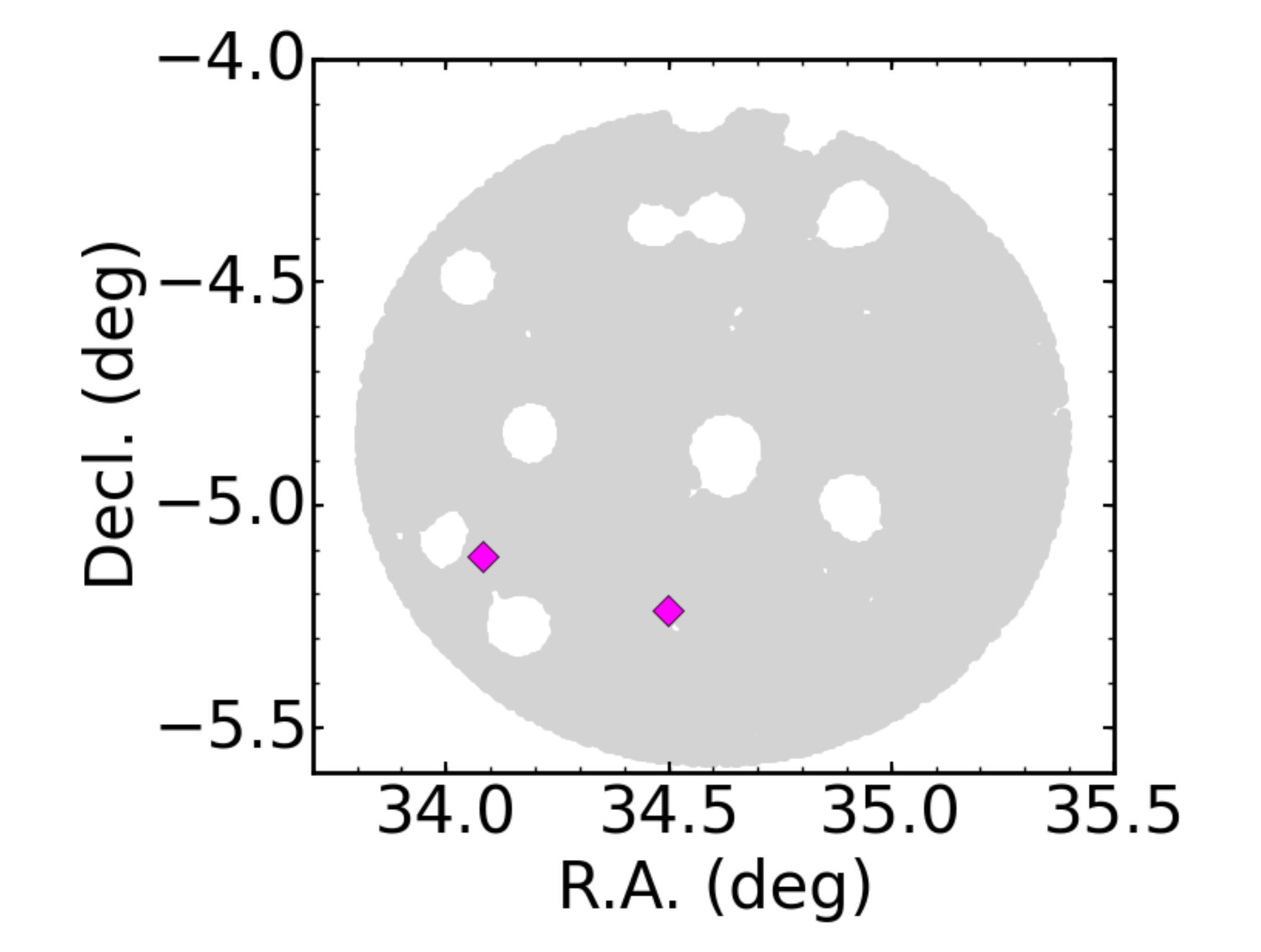}
\caption{
Top: Sky distribution of the LAE candidates in the COSMOS field.
The red squares, the magenta diamonds and the black circles represent positions of 
bright ($\log L_{{\rm Ly\alpha}}[{\rm erg\ s^{-1}}]>43.3$),
medium-bright ($\log L_{{\rm Ly\alpha}}[{\rm erg\ s^{-1}}]=43.1-43.3$),
and faint ($\log L_{{\rm Ly\alpha}}[{\rm erg\ s^{-1}}]<43.1$) LAEs, respectively.
The red open squares represent the LAE-1 and LAE-3 that are also found by \cite{zhengy2017}
and spectroscopically confirmed as $z=6.936$ and $z=6.931$ LAEs by \cite{hu2017}.
Bottom: 
Same as the top panel, but for the LAE candidates in the SXDS field.
\label{fig:skydist}}
\end{minipage}
\end{figure}

In this way, we define the selection criteria of $z=7.0$ LAEs:
\begin{eqnarray}
&&NB973_{\rm HSC} < NB973_{\rm HSC, 5\sigma} \nonumber \\
&&{\rm and}\ y - NB973_{\rm HSC} > 0.7 \nonumber \\
&&{\rm and}\ \left[ (z < z_{3\sigma}\ {\rm and}\ z - y > 2.0)\ {\rm or}\ z > z_{3\sigma} \right] \\
&&{\rm and}\ g > g_{3\sigma}\ {\rm and}\ r>r_{3\sigma}\ {\rm and}\ i>i_{3\sigma} \nonumber \\
&&{\rm and}\ NB816 > NB816_{3\sigma}\ {\rm and}\ NB921 > NB921_{3\sigma}\ \nonumber \\
&&{\rm and}\ NB718 > NB718_{3\sigma}, \nonumber
\label{eq:selection_criteria}
\end{eqnarray}
where the indices of $5\sigma$ and $3\sigma$ denote the $5\sigma$ and $3\sigma$ detection limits of the images, respectively. 
We use $2''.0$-diameter aperture magnitudes to measure the S/N values for source detections, 
and {\tt cmodel} magnitudes for color measurements.
In addition to these color selection criteria, we use the {\tt countinputs} parameter generated by {\tt hscPipe}, 
which indicates the number of stacked image frames for each object in each band.  
We apply ${\tt countinputs} \geq 3$ for the $NB973_{\rm HSC}$ images. 
We also use the following flags of {\tt hscPipe}: 
{\tt flags} {\tt pixel} {\tt edge}, {\tt flags} {\tt pixel} {\tt interpolated} {\tt center}, 
{\tt flags} {\tt pixel} {\tt saturated} {\tt center}, {\tt flags} {\tt pixel} {\tt cr} {\tt center}, and {\tt flags} {\tt pixel} {\tt bad}, 
to remove objects with 
bad pixels 
or a poor photometric measurement 
(see \citealt{shibuya2018a} for more details).
Then we perform visual inspections for $NB$ and $BB$ images of all the objects
which pass the selection criteria to exclude objects affected by cosmic rays, cross-talk, and diffuse halo near bright stars. 
Although we impose the criteria of no detection more than $3\sigma$ detection level in these bands (e.g., $g>g_{3\sigma}$), 
we also remove objects
which have possible counterparts in $g,\ r,\ i$, $NB718$, $NB816$, or $NB921$ bands. 
 After the visual inspection, 32 and 2 LAE candidates are selected 
in COSMOS and SXDS fields, respectively (Table \ref{tab:LAE_summary}).
We show the spatial distribution of our LAEs in Figure \ref{fig:skydist}.

We compare our $z=7$ LAE sample with those obtained by the previous studies \citep{ota2017, zhengy2017}. 
\cite{ota2017} identify 6 LAE candidates in the SXDS field.
We select two LAE candidates, 
HSC-z7LAE33 and HSC-z7LAE34, in the SXDS field.  
HSC-z7LAE33 is also selected by \cite{ota2017} (NB973-SXDS-S-95993 in their paper) in the SXDS field. 
HSC-z7LAE34 is not identified in \cite{ota2017}, 
because it is located in the SXDS 
field outside the \cite{ota2017} observation footprints. 
We do not identify the other LAEs selected by \cite{ota2017}, 
because the limiting magnitude of our $NB973_{\rm HSC}$ images in the SXDS field is shallower than that of Ota et al.'s $NB973_{\rm SC}$ images. 

In the COSMOS field, HSC-z7LAE3 and HSC-z7LAE25 are previously selected by \cite{zhengy2017} as LAE-1 and LAE-3, respectively. 
HSC-z7LAE3 and HSC-z7LAE25 are spectroscopically confirmed by \cite{hu2017} using IMACS on Magellan.
According to \cite{hu2017}, 
the redshifts of HSC-z7LAE3 and HSC-z7LAE25 are $z=6.936$ and 6.931, respectively. 
Because the filter responses of Zheng et al.'s $NB964$ and our $NB973_{\rm HSC}$ are different, 
we do not identify the LAEs selected by \cite{zhengy2017} 
except for the luminous LAEs, HSC-z7LAE3 and HSC-z7LAE25. 

\begin{deluxetable*}{ccccccccccc}
\tablecolumns{7}
\tabletypesize{\scriptsize}
\tablecaption{Photometry of the z=7 LAE candidates
\label{tab:LAE_summary}}
\setlength{\tabcolsep}{0.0in}
\tablewidth{0pt}
\tablehead{
\colhead{ID} & 
\colhead{R.A.} & 
\colhead{Decl.} & 
\colhead{$y_{\rm total}$} &
\colhead{$NB973_{\rm HSC}$} &
\colhead{$NB973_{\rm HSC, total}$} &
\colhead{$L_{\rm Ly\alpha}$}\\
\colhead{(1)} & 
\colhead{(2)} & 
\colhead{(3)} & 
\colhead{(4)} &
\colhead{(5)} &
\colhead{(6)} &
\colhead{(7)}
}
\startdata
HSC-z7LAE1&10:02:15.5&$+$02:40:33.4&25.09&23.52&23.40&29.81\\
HSC-z7LAE2&10:02:23.4&$+$02:05:05.1&25.63&23.92&23.68&24.12\\
HSC-z7LAE3\tablenotemark{a}&10:02:06.0&$+$ 02:06:46.2&25.04&24.09&23.77&39.32\tablenotemark{b}\\
HSC-z7LAE4&10:01:41.9&$+$01:40:03.6&25.59&24.51&24.10&14.86\\
HSC-z7LAE5&10:00:20.3&$+$02:20:04.2&26.40&24.31&24.11&16.94\\
HSC-z7LAE6&10:03:04.4&$+$02:17:15.1&25.69&24.38&24.12&14.81\\
HSC-z7LAE7&10:01:55.9&$+$02:50:33.6&26.38&24.32&24.20&15.33\\
HSC-z7LAE8&09:59:27.6&$+$01:41:01.3&25.37&24.46&24.25&11.46\\
HSC-z7LAE9&10:01:01.4&$+$02:33:51.2&26.16&24.46&24.28&13.68\\
HSC-z7LAE10&10:01:16.9&$+$02:21:04.2&26.28&24.52&24.29&13.75\\
HSC-z7LAE11&10:02:25.3&$+$01:59:23.2&$>26.85$&24.8&24.37&13.32\\
HSC-z7LAE12&09:59:00.7&$+$02:14:18.4&$>26.85$&24.54&24.39&12.99\\
HSC-z7LAE13&09:57:59.4&$+$02:36:32.4&$>26.85$&24.76&24.40&12.86\\
HSC-z7LAE14&10:01:32.9&$+$02:41:55.6&26.07&24.67&24.42&11.54\\
HSC-z7LAE15&10:01:59.4&$+$02:29:30.4&26.40&24.56&24.44&12.01\\
HSC-z7LAE16&10:02:56.5&$+$02:17:22.6&$>26.85$&24.62&24.44&8.33\\    
HSC-z7LAE17&10:00:12.9&$+$02:30:47.1&26.02&24.72&24.46&10.81\\
HSC-z7LAE18&09:58:38.3&$+$01:47:49.6&$>26.85$&24.86&24.47&12.04\\
HSC-z7LAE19&09:59:58.7&$+$01:30:33.4&26.06&24.7&24.49&10.60\\
HSC-z7LAE20&10:02:12.0&$+$02:47:40.6&25.76&24.63&24.51&9.42\\
HSC-z7LAE21&09:57:49.1&$+$02:34:36.4&$>26.85$&24.84&24.52&11.39\\
HSC-z7LAE22&10:02:47.1&$+$02:10:40.1&26.84&24.80&24.52&11.35\\
HSC-z7LAE23&10:01:04.5&$+$02:12:09.2&26.51&24.85&24.53&11.09\\
HSC-z7LAE24&10:02:37.8&$+$02:13:39.2&26.50&24.79&24.56&10.64\\
HSC-z7LAE25\tablenotemark{c}&10:01:53.5&$+$ 02:04:59.6&25.74&24.96&24.75&24.55\tablenotemark{d}\\    
HSC-z7LAE26&10:00:26.0&$+$02:31:39.0&26.56&24.77&24.62&10.08\\
HSC-z7LAE27&09:59:17.1&$+$02:47:02.5&26.10&24.95&24.62&9.12\\
HSC-z7LAE28&09:59:36.4&$+$02:06:05.5&26.78&24.91&24.68&9.63\\
HSC-z7LAE29&10:00:39.2&$+$02:04:56.9&26.02&24.93&24.68&8.36\\
HSC-z7LAE30&09:59:52.6&$+$02:40:01.8&$>26.85$&24.84&24.71&9.27\\
HSC-z7LAE31&10:02:39.4&$+$02:07:12.1&26.71&24.86&24.78&8.60\\
HSC-z7LAE32&10:00:37.4&$+$02:43:14.7&$>26.85$&24.94&24.85&7.97\\ \hline
HSC-z7LAE33\tablenotemark{e}&02:17:59.5&$-$05:14:07.43&25.47&24.26&24.00&18.00\\
HSC-z7LAE34&02:16:20.1&$-$05:07:01.2&$>26.65$&24.39&24.16&16.75
\enddata
\tablecomments{
(1): Object ID. (2)-(3): RA and Dec. (4): The {\tt cmodel} magnitudes in $y$ band. 
The lower limit corresponds to a $1\sigma$ limit.
(5): The $2''$-aperture magnitudes in $NB973_{\rm HSC}$.
(6): The {\tt cmodel} magnitudes in $NB973_{\rm HSC}$. 
(7): The Ly$\alpha$ luminosities in $10^{42}$ erg s$^{-1}$.
}
\tablenotetext{a}{
HSC-z7LAE3 is the LAE that is also identified by \cite{zhengy2017} (LAE-1 in their paper). 
This object is previously spectroscopically confirmed as the $z=6.936$ LAE by \cite{hu2017}.
}
\tablenotetext{b}{
We estimate the Ly$\alpha$ luminosity of HSC-z7LAE3, assuming that the redshift is $z=6.936$ \citep{hu2017}. 
}
\tablenotetext{c}{
HSC-z7LAE25 is the LAE that is also identified by \cite{zhengy2017} (LAE-3 in their paper). 
This object is previously spectroscopically confirmed as the $z=6.931$ LAE by \cite{hu2017}.
}
\tablenotetext{d}{
We estimate the Ly$\alpha$ luminosity of HSC-z7LAE25, assuming that the redshift is $z=6.931$ \citep{hu2017}. 
}
\tablenotetext{e}{
HSC-z7LAE33 is the LAE candidate that is also selected by \cite{ota2017} (NB973-SXDS-S-95993 in their paper). 
}
\end{deluxetable*}

\section{Luminosity Function}
\label{sec:luminosity_function}

\subsection{Detection Completeness and Surface Number Density}
\label{sec:photometric_lae_completeness}
We estimate the detection completeness of our $NB973_{\rm HSC}$ images using Monte Carlo simulations described in \citet{konno2018}
with the {\tt SynPipe} software \citep{huang2017, murata2017}.
Using the {\tt SynPipe} software, we distribute $\sim24,000$ pseudo LAEs
with various magnitudes in each $NB973_{\rm HSC}$ frame in each field.
We then stack the image frames, and detect these input LAEs with {\tt hscPipe}. 
These pseudo LAEs have a S\'{e}rsic index of $n=1.5$ 
and a half-light radius of $r_e\sim0.8\ {\rm kpc}$.
These values are similar to those of $z\sim7$ Lyman break galaxies (LBGs) with $L_{{\rm UV}}=0.3-1L_{z=3}^*$ \citep{shibuya2015}. 
Our HSC data are too shallow ($\sim 10^{-18}\ {\rm erg}^{-1}\ {\rm s}^{-1}\ {\rm cm}^{-2}\ {\rm arcsec}^{-2}$) to identify the extended Ly$\alpha$ halo.
One needs data deeper than our HSC data by an order of magnitude to detect the extended Ly$\alpha$ halo ($\sim 10^{-19}\ {\rm erg}^{-1}\ {\rm s}^{-1}\ {\rm cm}^{-2}\ {\rm arcsec}^{-2}$).
Our HSC data thus detect the central core component of an LAE.
The half-light radius of our pseudo LAEs is consistent with 
that of the Ly$\alpha$ emission from the core component 
obtained by the recent MUSE spectroscopic survey \citep{leclercq2017}.

\begin{figure}
\epsscale{1.25}
\plotone{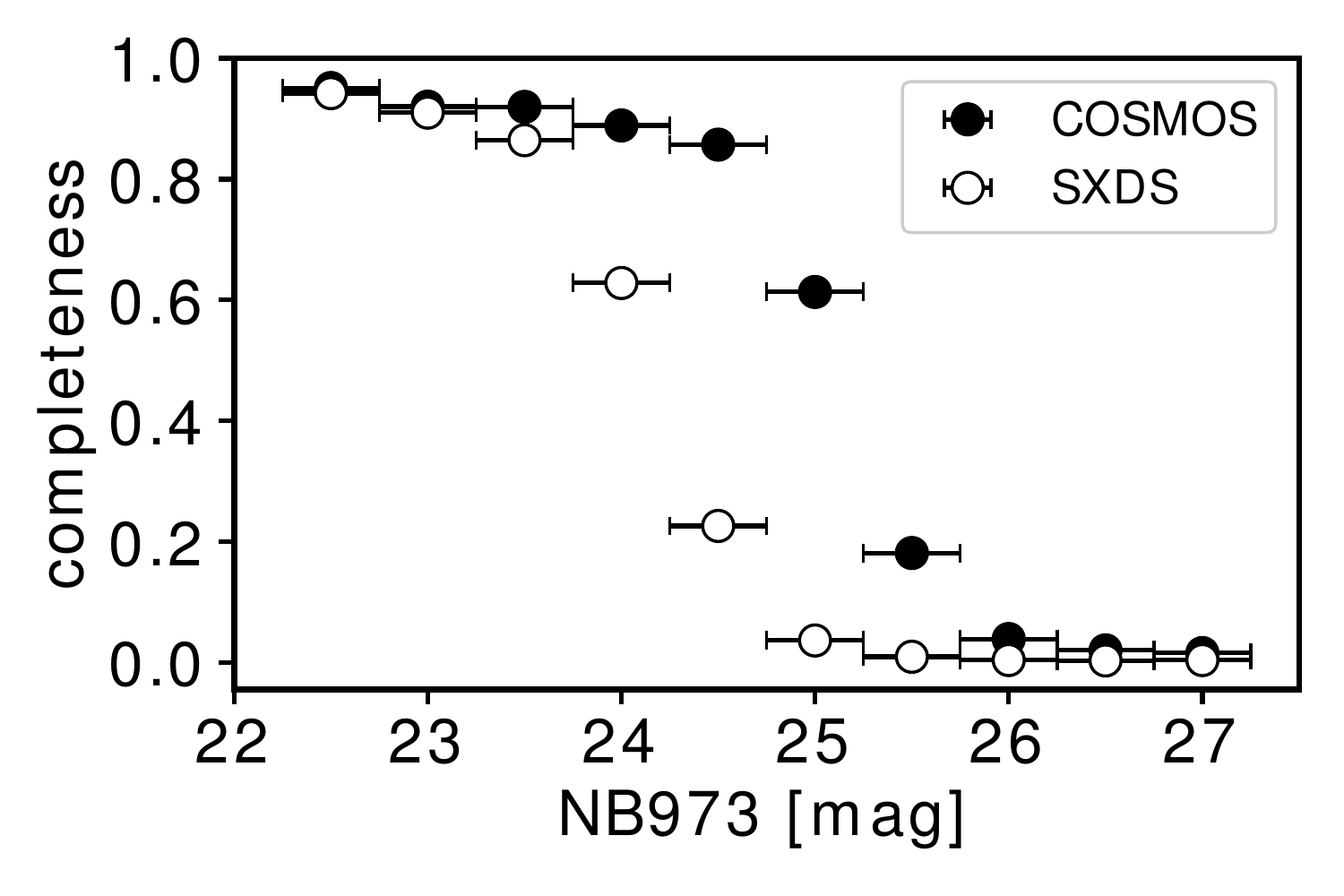}
\caption{
Detection completeness of our $NB973_{\rm HSC}$ images as a function of $NB973_{\rm HSC}$ magnitude. 
The 5$\sigma$ limiting magnitudes of our $NB973_{\rm HSC}$ images are 
24.96 and 24.26 in COSMOS and SXDS fields, respectively.
\label{fig:completeness}}
\end{figure}

We define the detection completeness as a fraction of the number of detected pseudo LAEs to all of the input pseudo LAEs. 
We show the detection completeness in Figure \ref{fig:completeness}. 
We find that the detection completeness is $\gtrsim80-90$\% for relatively luminous sources 
($\lesssim24.5$ and $23.5$ mag in COSMOS and SXDS fields, respectively), 
and $\sim60$\% at the $5\sigma$ limiting magnitudes in each field.

\begin{figure}
\epsscale{1.25}
\plotone{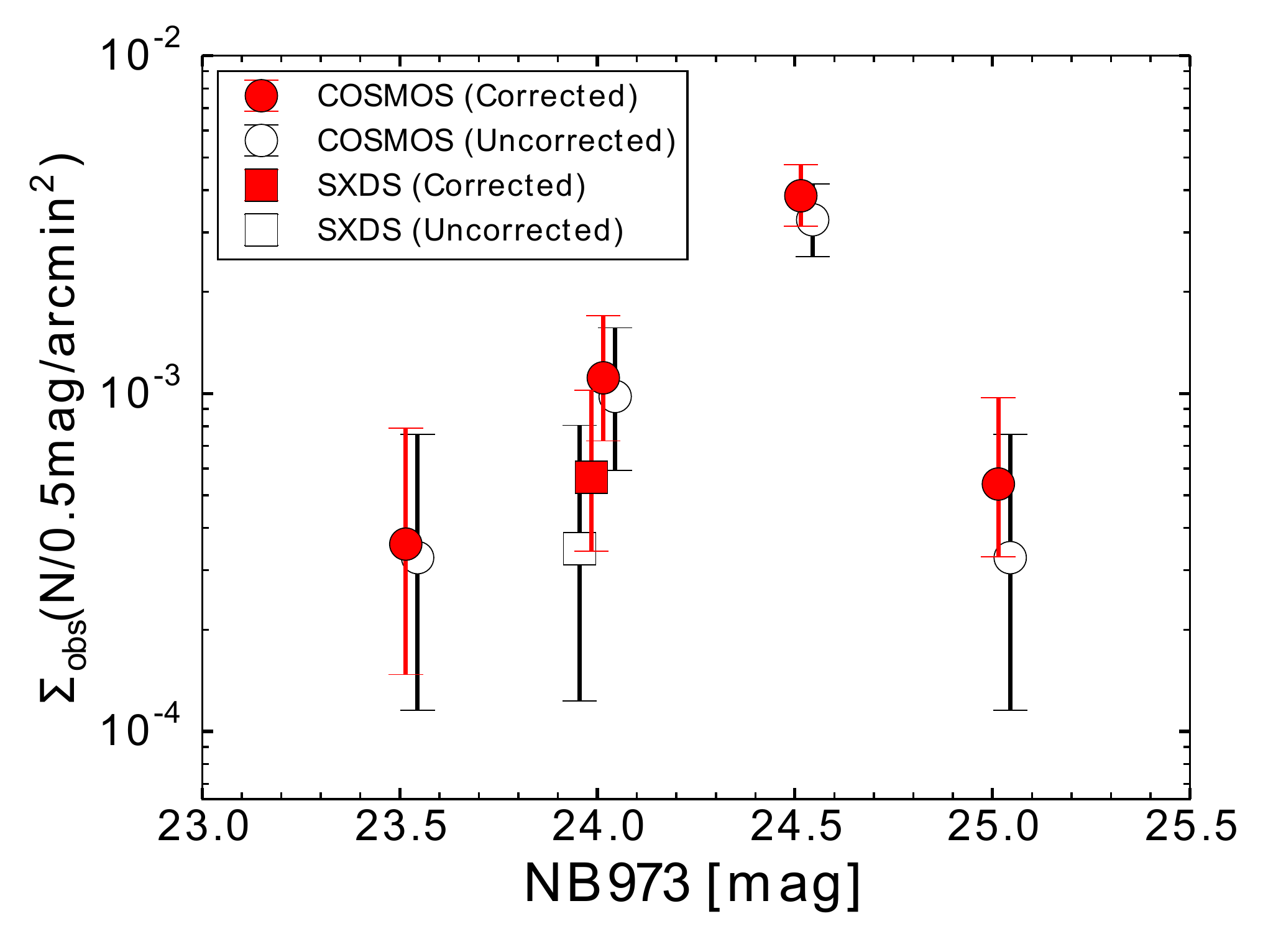}
\caption{
Surface number densities of our $z=7.0$ LAEs as a function of $NB973_{\rm HSC}$ magnitude.
The red filled circles and square represent the surface number densities in COSMOS and SXDS fields, respectively, with the completeness correction.
The black open circles and square are the surface number densities with no completeness correction.
\label{fig:surface_number_density}}
\end{figure}

We derive the surface number densities as a function of $NB973_{\rm HSC}$ magnitude. 
The surface number density is defined by 
the number of the sources in each magnitude bin 
divided by the survey area and the detection completeness. 
We show the surface number densities in Figure \ref{fig:surface_number_density}.
The errors of the surface number density are calculated 
based on the Poisson errors for the small number statistics \citep{gehrels1986}.
We use the values in the columns ``0.8413" in Tables 1 and 2 of \cite{gehrels1986}
for $1\sigma$ upper and lower confidence intervals, respectively.

\subsection{Ly$\alpha$ Luminosity Function}
\label{sec:ly_alpha_luminosity_function}

We derive the Ly$\alpha$ LF in the same manner as \citet{ouchi2008} and \citet{ouchi2010}. 
We obtain the volume number density of LAEs in a Ly$\alpha$ luminosity bin $[L,L+dL]$ with an equation defined by 

\begin{eqnarray}
\phi (L) dL = \sum_{i} \frac{1}{V_{\rm eff}\ f_{\rm comp}(m_{NB, i})} dL,
\label{eq:lf}
\end{eqnarray}
where the sum is taken over all objects $i$ in the luminosity bin. 
Here, $V_{\rm eff}$ is the survey volume estimated in Section \ref{sec:data_reduction}, 
and $f_{\rm comp}(m_{NB, i}$) is the detection completeness for an object $i$ with an $NB973$ magnitude of $m_{NB, i}$.
The bin size is 0.2 dex, which is the same as that of \cite{ota2017}.

We calculate the Ly$\alpha$ line flux ($f_{{\rm line}}$), 
and the rest-UV continuum flux ($f_{{\rm c}}$),
of each object 
from $NB$ and $BB$ magnitudes ($m_{NB}$ and $m_{BB}$) using the following formula 
\begin{eqnarray}
&&m_{NB, BB} + 48.6 = -2.5 \log \frac{\int ^{\nu_{{\rm Ly\alpha}}}_{0} \left(f_c + f_{{\rm line}}\right)T_{NB, BB} d\nu/\nu}{\int T_{NB, BB} d\nu/\nu}.\nonumber \\
&&\ 
\label{eq:lyaflux}
\end{eqnarray}
Here, $T_{NB}$ and $T_{BB}$ are the transmission curves of the $NB$ and $BB$ filters, respectively.
We use $NB973_{\rm HSC}$ and $y$-band {\tt cmodel} magnitudes for $m_{NB}$ and $m_{BB}$, respectively.
$\nu_{{\rm Ly\alpha}}$ is the observed frequency of the Ly$\alpha$ line.
Because HSC-z7LAE3 and HSC-z7LAE25 have the spectroscopic redshifts of $z=6.936$ and $6.931$, 
we use $\nu_{{\rm Ly\alpha}}=3.108\times10^{14}$ Hz and $3.110\times10^{14}$ Hz for the calculation of the Ly$\alpha$ line fluxes, respectively.
In the calculation of the Ly$\alpha$ line flux of the other LAEs, 
we adopt $\nu_{{\rm Ly\alpha}}=3.087\times10^{14}$ Hz 
corresponding to the central frequency of our $NB973_{\rm HSC}$ bandpass.
We assume that $f_{{\rm line}}$ is a $\delta$-function, and that $f_c$ is a constant.
We also assume that the flux bluewards of Ly$\alpha$ is zero due to the IGM absorption.
If an LAE is not detected in $BB$, we replace $m_{BB}$ by the $1\sigma$ limiting magnitudes of $BB$.
We set $f_c$ to 0 if the condition of $f_c < 0$ is met.

We include uncertainties from the Poisson statistics, the cosmic variance and the contamination rate for the error bars of each bin. 
Again, we apply the result from \cite{gehrels1986} for the Poisson errors.
For the cosmic variance $\sigma_{\rm g}$ estimate, 
we use the relation
\begin{eqnarray}
\sigma_{\rm g} = b_{\rm g} \sigma_{{\rm DM}} (z, R), 
\label{eq:cosmicvariance}
\end{eqnarray}
where $b_{\rm g}$ and $\sigma_{{\rm DM}}(z, R)$ are the bias and the density fluctuation
of dark matter, respectively, at the redshift of $z$ in a radius of $R$.
\cite{ouchi2018} derive the bias parameter of $b_g=4.5\pm0.6$ at $z=6.6$
from the sample of 873 LAEs in a total of $21.2\ {\rm deg^2}$ area including
COSMOS and SXDS fields.
Here we adopt $b_{\rm g} = 4.5$ for the $z=7.0$ LAEs, assuming that $b_{\rm g}$ does not evolve
significantly at $z=6.6-7.0$.
We obtain $\sigma (z, R)=0.038$ at $z=7.0$ using the analytic cold dark matter model
\citep{sheth1999, Mo2002} and our survey volumes in COSMOS and SXDS fields.
With this procedure, we estimate the fractional uncertainty from the cosmic variance to be $\sigma_g = 0.17$ for a one-field Ly$\alpha$ LF and $\sigma_g = 0.12$ for the total Ly$\alpha$ LF. 

\begin{figure}
\epsscale{1.25}
\plotone{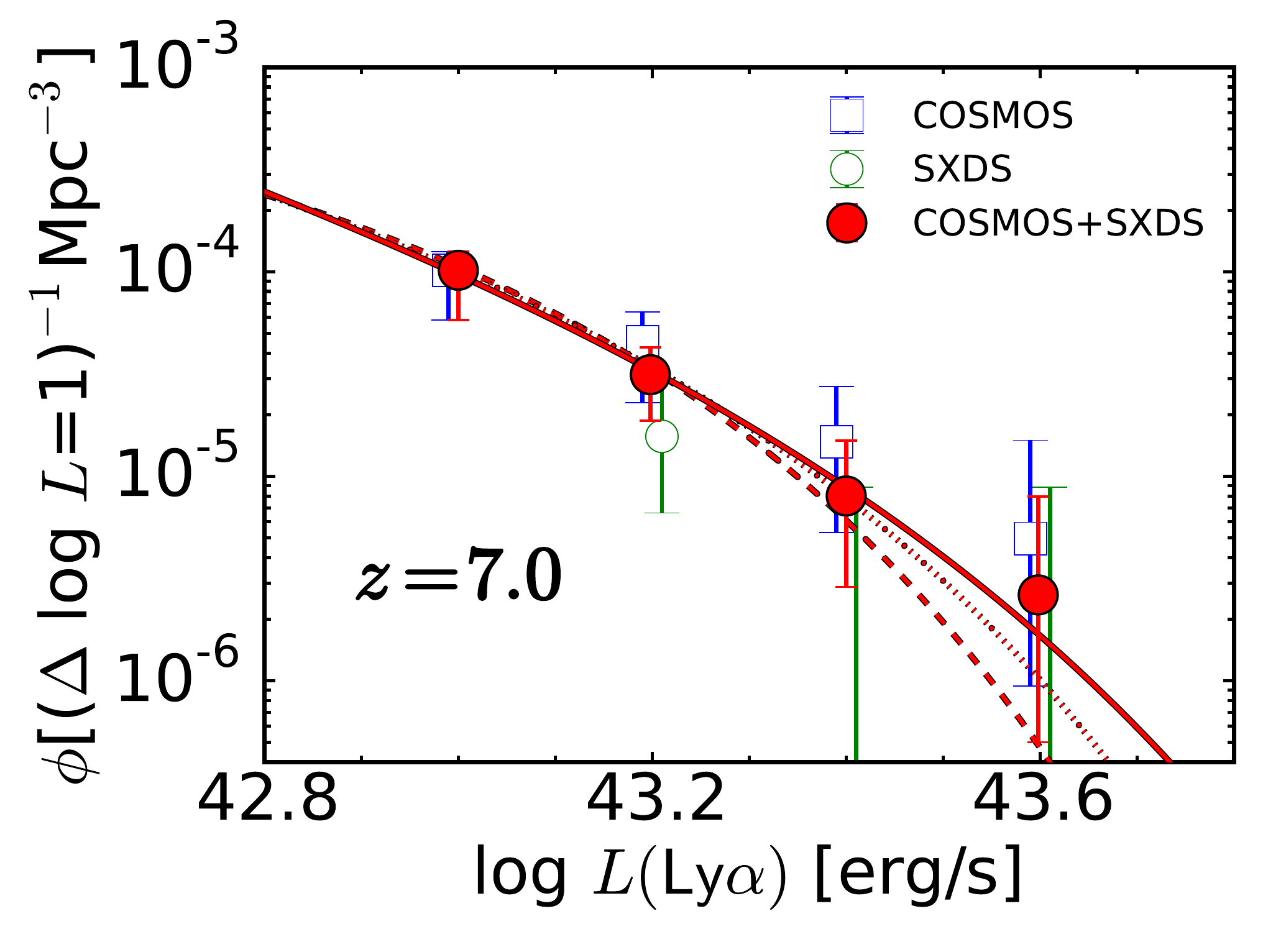}
\caption{
Ly$\alpha$ LFs of our $z=7.0$ LAEs.
The red filled circles represent our best estimate of 
the Ly$\alpha$ LF derived with the data from both COSMOS and SXDS fields.
The best-fit Schechter function for the entire fields is shown with the red solid, dotted, and dashed curves 
with a fixed faint-end slope of $\alpha=-2.5,\ -2.0$, and $-1.5$, respectively.
The blue open squares and green open circle denote our Ly$\alpha$ LFs estimated with
the data of the COSMOS and SXDS fields, respectively. In two bright luminosity bins in the SXDS field,
we also plot the $1\sigma$ upper error of the Ly$\alpha$ LF.
\label{fig:LF1}}
\end{figure}

Because our sample consists of the $z=7$ photometric LAE candidates except for one LAE spectroscopically confirmed \citep{hu2017}, 
we do not determine the contamination rate with our sample. 
To assess the contamination rate of our sample, 
we refer the previous studies of narrowband surveys of LAEs. 
The contamination rate of $f_{{\rm cont}}=0-30\%$ is obtained in \citet{ouchi2008} and \citet{kashikawa2011},
who have conducted the Subaru/Suprime-Cam imaging survey for LAEs at $z =5.7$ and $6.6$.
\cite{shibuya2018b} have conducted the spectroscopic follow-up observations for 
$z=5.7$ and $6.6$ LAE candidates obtained in the HSC survey \citep{konno2018}.
They confirm 13 sources out of 18 candidates, and derived the contamination rate of $f_{{\rm cont}}\simeq30\%$.
We take into account the uncertainty of the contamination by increasing the lower 1$\sigma$ confidence intervals of the Ly$\alpha$ LF by 30\%.
Figure \ref{fig:LF1} represents the LF of our $z=7.0$ LAEs.
The Ly$\alpha$ LFs of COSMOS and SXDS fields are consistent within the uncertainties.

\begin{figure}
\epsscale{1.25}
\plotone{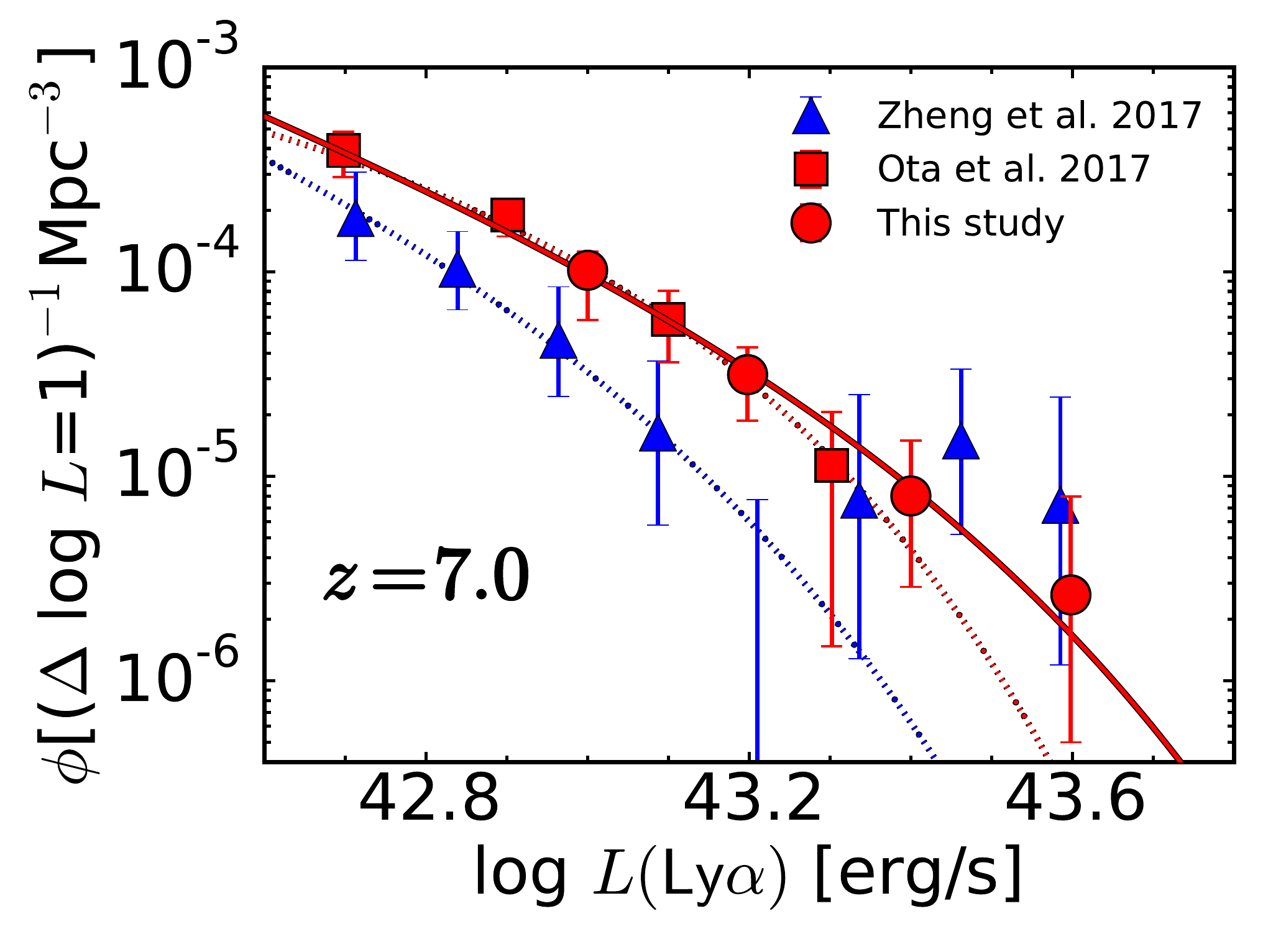}
\caption{
Comparison of our Ly$\alpha$ LF of LAEs at $z\simeq7.0$ 
with those derived by previous studies.
The red filled circles denote our results at $z=7.0$ 
and the red solid curve is the best-fit Schechter function 
with $\alpha=-2.5$.
The red squares and the dotted curve represent the Ly$\alpha$ LF at $z=7.0$
and the best-fit Schechter function with the fixed faint end slope of $\alpha=-1.5$ given by \cite{ota2017}.
The blue triangles and the solid curve represent the Ly$\alpha$ LF at $z=6.9$
and the best-fit Schechter function over the luminosity range of $\log L_{{\rm Ly\alpha}}[{\rm erg\ s^{-1}}] =42.65-43.25$ given by \cite{zhengy2017}.
\label{fig:LF2}}
\end{figure}

In Figure \ref{fig:LF2}, we compare our $z=7.0$ Ly$\alpha$ LF with those obtained by previous studies.
We plot the Ly$\alpha$ LF at $z=7.0$ ($6.9$) derived by the Subaru Suprime-Cam \citep{ota2017} observations (DECam observations; \citealt{zhengy2017}).
The result of our study is consistent with those of \cite{ota2017} 
over a Ly$\alpha$ luminosity range of $\log L_{{\rm Ly\alpha}}[{\rm erg\ s^{-1}}] \sim42.9-43.3$.
At the bright end, $\log L_{{\rm Ly\alpha}}[{\rm erg\ s^{-1}}] > 43.3$, our Ly$\alpha$ LF is consistent with that of \cite{zhengy2017}. 
On the other hand, measurements of \cite{zhengy2017} fall below our data points 
at the relatively faint end, $\log L_{{\rm Ly\alpha}}[{\rm erg\ s^{-1}}]<43.3$.
This difference between our and Zheng et al.'s results 
may be caused by the systematic uncertainty of the completeness correction 
of the faint end (Z. Zheng,  private communication).

We fit a Schechter function \citep{schechter1976} to our $z=7.0$ Ly$\alpha$ LF
by minimum $\chi^2$ fitting. The Schechter function is defined by
\begin{eqnarray}
\phi(L) && d\log L \nonumber \\
&&= \ln 10\ \phi^* \left(\frac{L}{L^*}\right)^{\alpha+1} \exp \left(-\frac{L}{L^*}\right) d \log L,
\label{eq:schechter}
\end{eqnarray}
where $L^*$ and $\phi^*$ represent the characteristic luminosity and number density, respectively,
and $\alpha$ is the faint-end slope.

We determine the best-fit values of $\phi^*$ and $L^*$ for a series of possible values of $\alpha$.
We include the faint-end Ly$\alpha$ LF of \cite{ota2017} that is consistent with our results, 
and cover the faint Ly$\alpha$ luminosity range that we do not reach.
Specifically, we use two faint-end data points of \cite{ota2017} in the luminosity range of $\log L_{{\rm Ly\alpha}}[{\rm erg\ s^{-1}}]=42.6-43.0$. 
In this luminosity range, 
we confirm that there is no overlap of the LAEs 
selected by Ota et al.'s and our studies. 
Note that we do not use the two bright-end data points of \cite{ota2017} 
in the luminosity range of $\log L_{{\rm Ly\alpha}}[{\rm erg\ s^{-1}}]=43.0-43.3$, 
because they are not statistically independent 
of our data points. 
Because the difference in $\chi^2$ for $\alpha$ values is insignificant,
we fix the faint-end slope to $\alpha=-2.5$, $-2.0$, and $-1.5$.
We use six luminosity bins in total for the fitting.
The number of the bins for the fitting of the Schechter function is 
comparable to those of previous Ly$\alpha$ LF studies 
(e.g., \citealt{ouchi2010,matthee2015,santos2016}). 
The best-fit Schechter parameters are summarized in Table \ref{tab:schechter_param}. 
Figure \ref{fig:LF1} shows the best-fit Schechter functions 
with the red dashed, dotted, and solid lines for $\alpha=-1.5$, $-2.0$, and $-2.5$, respectively.
The best-fit Schechter functions are consistent with the bright end of our Ly$\alpha$ LF within the error bar, 
for any faint-end slopes steeper than $\alpha=-1.5$. 
The previous HSC LAE study obtains the very steep faint-end slope of $\alpha=-2.5$ 
for the Ly$\alpha$ LFs at $z=5.7$ and 6.6 \citep{konno2018}.
Moreover, similar values of $\alpha$ are also reported 
by the MUSE spectroscopic survey for LAEs at $z=3-6.6$ that reaches a Ly$\alpha$ luminosity 
as faint as $\log L_{{\rm Ly\alpha}}[{\rm erg\ s^{-1}}]=41.5$ \citep{drake2017b}.
We adopt $\alpha=-2.5$ as our fiducial value. 
We find no clear signature of bright-end excess over the best-fit Schechter function 
that is claimed by \cite{zhengy2017} (see Figure \ref{fig:LF2}). 

\begin{figure}
\epsscale{1.25}
\plotone{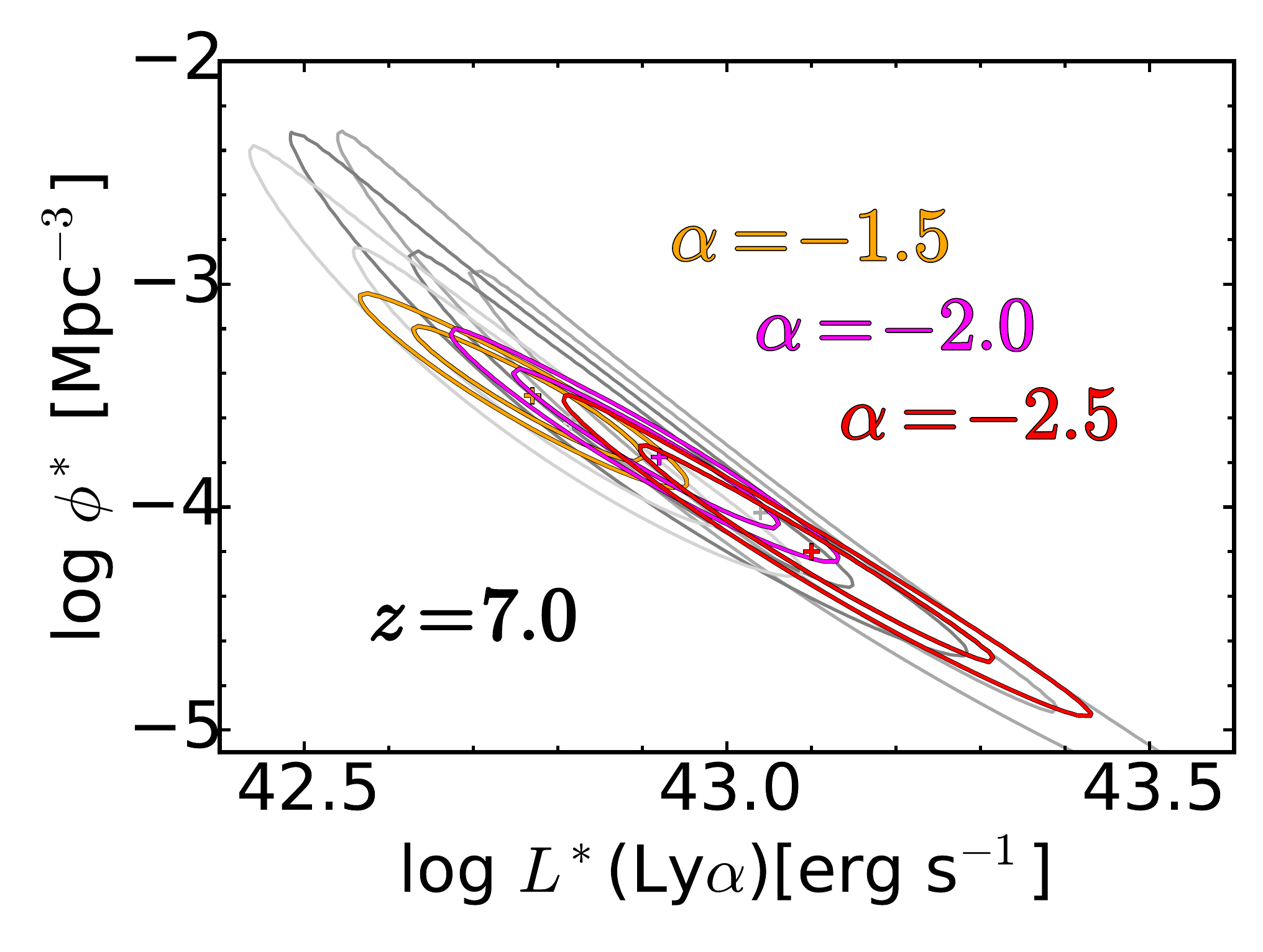}
\caption{
Confidence intervals of the Schechter parameters 
for the Ly$\alpha$ LF at $z=7.0$. 
All of the inner and outer contours correspond to 
68\% and 90\% confidence intervals, respectively. 
The red (dark-gray), magenta (gray), and orange (light-gray) contours represent the fit to our $z=7.0$ LF 
for $\alpha=-2.5$, $-2.0$, and $-1.5$, respectively, 
with (without) the two data points in \cite{ota2017}. 
\label{fig:contour1}}
\end{figure}

We obtain the error contours of the Schechter parameters for 
the $68\%$ and $90\%$ confidence levels 
using the minimum $\chi^2$ method (e.g., \citealt{avni1976}). 
We define the error contours of the $68\%$ and $90\%$ confidence levels 
as the Schechter parameters corresponding to  
$\Delta \chi^2=2.30$ and $4.61$, respectively. 
Here, $\Delta \chi^2$ is the difference between 
$\chi^2$ and the $\chi^2$ minimum $\chi^2_{\rm min}$ ($\Delta \chi^2 = \chi^2 - \chi^2_{\rm min}$).
Figure \ref{fig:contour1} shows the error contours of the Schechter parameters 
of the Ly$\alpha$ LFs at $z=7.0$. 
The red (dark-gray), magenta (gray), and orange (light-gray) contours 
represent the results of the fitting 
in the case of $\alpha=-2.5$, $-2.0$, and $-1.5$, respectively, 
with (without) the two faint-end data points of \cite{ota2017}. 
We find that most of the error contours overlap each other. 
However, the error contours for $\alpha=-2.5$ (red) and $-1.5$ (orange) 
barely overlap at the $68\%$ confidence level. 
This difference suggests that the best-fit result of 
$\alpha=-1.5$ is not as good as that of $\alpha=-2.5$. 
In fact, the Schechter function with 
$\alpha=-2.5$ is well fitted to our Ly$\alpha$ LF over the entire luminosity range, 
while the best-fit result for $\alpha=-1.5$ does not agree 
with the brightest data point 
falling above the error bar (see Figure \ref{fig:LF1}). 

\begin{deluxetable*}{ccccccccc}
\tablecolumns{6}
\tabletypesize{\scriptsize}
\tablecaption{Best-fit Schechter Parameters and Ly$\alpha$ LDs
\label{tab:schechter_param}}
\setlength{\tabcolsep}{0.0in}
\tablewidth{0pt}
\tablehead{
\colhead{Redshift} & 
\colhead{$L^*$} &
\colhead{$\phi^*$} & 
\colhead{$\alpha$} &
\colhead{$\rho^{{\rm Ly\alpha}}_{{\rm obs}}$\tablenotemark{a}} &
\colhead{Reference} \\
\colhead{} & 
\colhead{ ($10^{43}\ {\rm erg\ s^{-1}}$)} &
\colhead{($10^{-4}\ {\rm Mpc^{-3}}$) } & 
\colhead{} &
\colhead{($10^{39}\ {\rm erg\ s^{-1}\ Mpc^{-3}}$)} &
\colhead{} 
}
\startdata
5.7      & $1.64^{+2.16}_{-0.62}$         & $0.849^{+1.87}_{-0.771}$     & $-2.56^{+0.53}_{-0.43}$ & $3.49^{+0.58}_{-0.71}$                 & \cite{konno2018} \\
6.6      & $1.66^{+0.30}_{-0.69}$         & $0.467^{+1.44}_{-0.442}$       & $-2.49^{+0.50}_{-0.50}$ & $1.82^{+0.30}_{-0.34}$                 & \cite{konno2018} \\
7.0      & $1.50^{+0.42}_{-0.31}$         & $0.45^{+0.26}_{-0.18}$       & $-2.5$ (fixed; fiducial)\tablenotemark{c}            & $1.43^{+0.45}_{-0.33}$                  & This study         \\
         & $0.92^{+0.38}_{-0.15}$        & $1.41^{+1.14}_{-0.77}$       & $-2.0$ (fixed)            & $1.27^{+0.25}_{-0.03}$                 & This study         \\
         & $0.63 ^{+0.11}_{-0.12}$        & $2.80^{+1.91}_{-0.90}$       & $-1.5$ (fixed)            & $1.16^{+0.13}_{-0.09}$                 & This study         \\
7.3      & $0.55^{+9.45}_{-0.33}  $    & $0.94^{+12.03}_{-0.93}$ & $-2.5$ (fixed; fiducial)\tablenotemark{c}         &             $0.34^{+0.58}_{-0.14}$     & \cite{konno2014}\tablenotemark{b}\\
         & $0.27^{+0.80}_{-0.12}$         & $3.7^{+17.6}_{-3.3}$         & $-1.5$ (fixed)            & $0.31^{+0.19}_{-0.12}$                 &  \cite{konno2014} 
\enddata
\tablenotetext{a}{
The Ly$\alpha$ LDs are obtained by integrating the Ly$\alpha$ LFs over the luminosity range of $\log L_{{\rm Ly\alpha}} = 42.4-44.0$.
}
\tablenotetext{b}{
The best-fit Schechter parameters are calculated by us using the data given by \cite{konno2014}. 
}
\tablenotetext{c}{
We choose $\alpha=-2.5$ as the fiducial value for the reason explained in Section \ref{sec:ly_alpha_luminosity_function}.
}
\end{deluxetable*}

\section{Discussion}
\label{sec:discussion}

\subsection{Evolution of Ly$\alpha$ Luminosity Functions at $z=5.7-7.3$}
\label{sec:evolution_of_lya_LF}

\begin{figure}
\epsscale{1.25}
\plotone{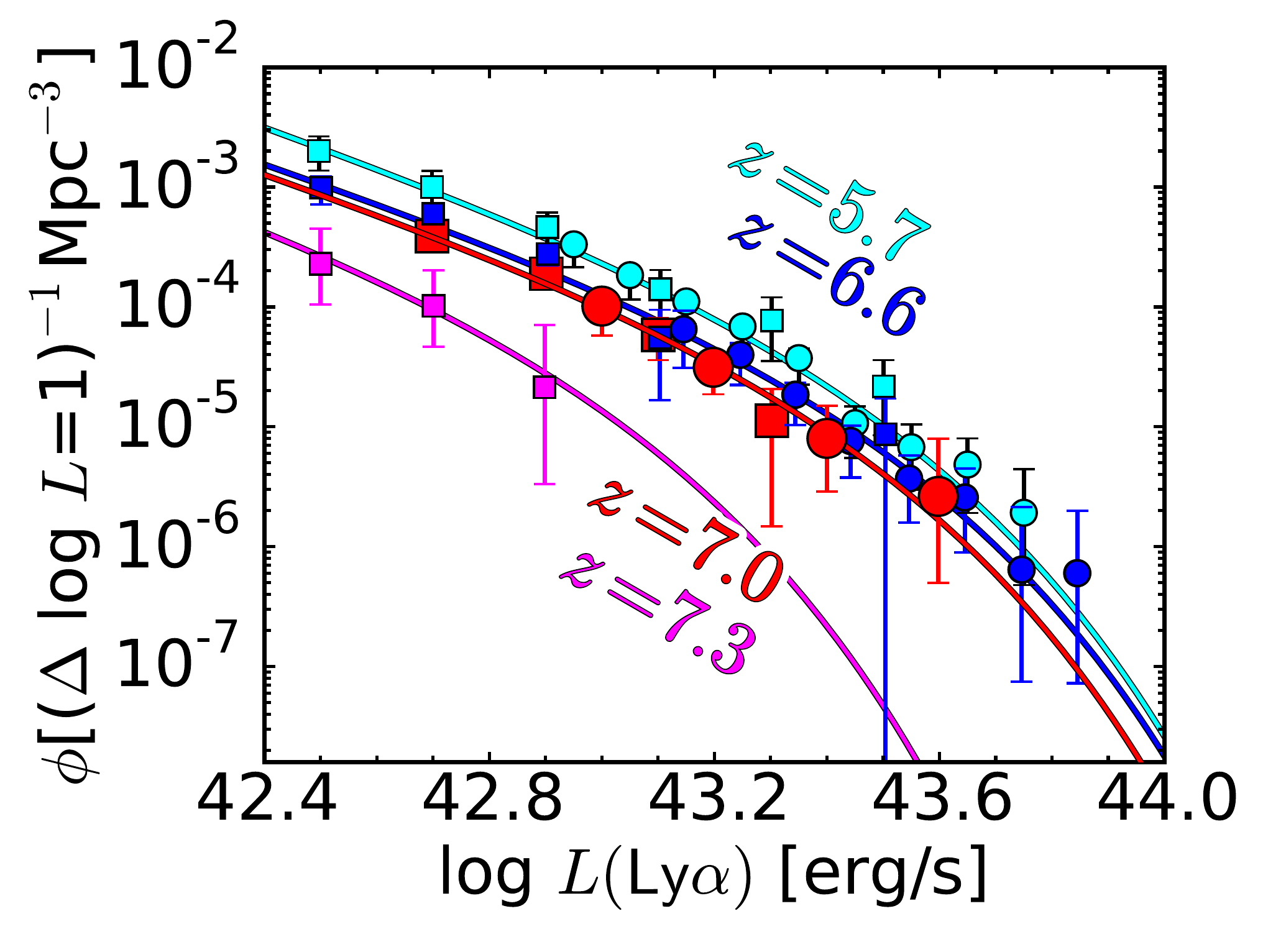}
\caption{
Evolution of Ly$\alpha$ LFs from $z=5.7$ to $7.3$.
The red filled circles are our Ly$\alpha$ LF at $z=7.0$ 
and the red solid line is the best-fit Schechter function.
The cyan and blue filled circles represent the $z=5.7$ and $6.6$  
Ly$\alpha$ LF measurements with the HSC data obtained by \cite{konno2018}.
The cyan, blue, red, and magenta filled squares represent the $z=5.7$, $6.6$, $7.0$, and $7.3$ 
Ly$\alpha$ LF measurements based on the Subaru/Suprime-Cam data 
derived by \cite{ouchi2008}, \cite{ouchi2010}, \cite{ota2017}, and \cite{konno2014}, respectively.
The cyan and blue solid curves are the best-fit Schechter functions for $z=5.7$ and $6.6$ Ly$\alpha$ LFs 
reported by \cite{konno2018}, respectively.
The magenta solid curve shows the best-fit Schechter function
to the Ly$\alpha$ LF at $z=7.3$ if the faint-end slope is fixed to $\alpha=-2.5$. 
\label{fig:LF3}}
\end{figure}

In Figure \ref{fig:LF3}, we plot our Ly$\alpha$ LF at $z=7.0$, 
and compare it with those at $z=5.7$, $6.6$, and $7.3$
derived by the previous Subaru LAE surveys \citep{ouchi2008, ouchi2010, konno2014, konno2018}.
At $z=7.0$ and 7.3, the solid lines indicate the best-fit Schechter functions with the fixed faint-end slope of $\alpha=-2.5$ 
for the reason explained in Section \ref{sec:ly_alpha_luminosity_function}. 
Our Ly$\alpha$ LF at $z=7.0$ shows a clear (small) decrease from the one at $z=5.7$ (6.6).  
The Ly$\alpha$ LF at $z=7.3$ displays a significant decrease from our Ly$\alpha$ LF at $z=7.0$.

\begin{figure}
\epsscale{1.25}
\plotone{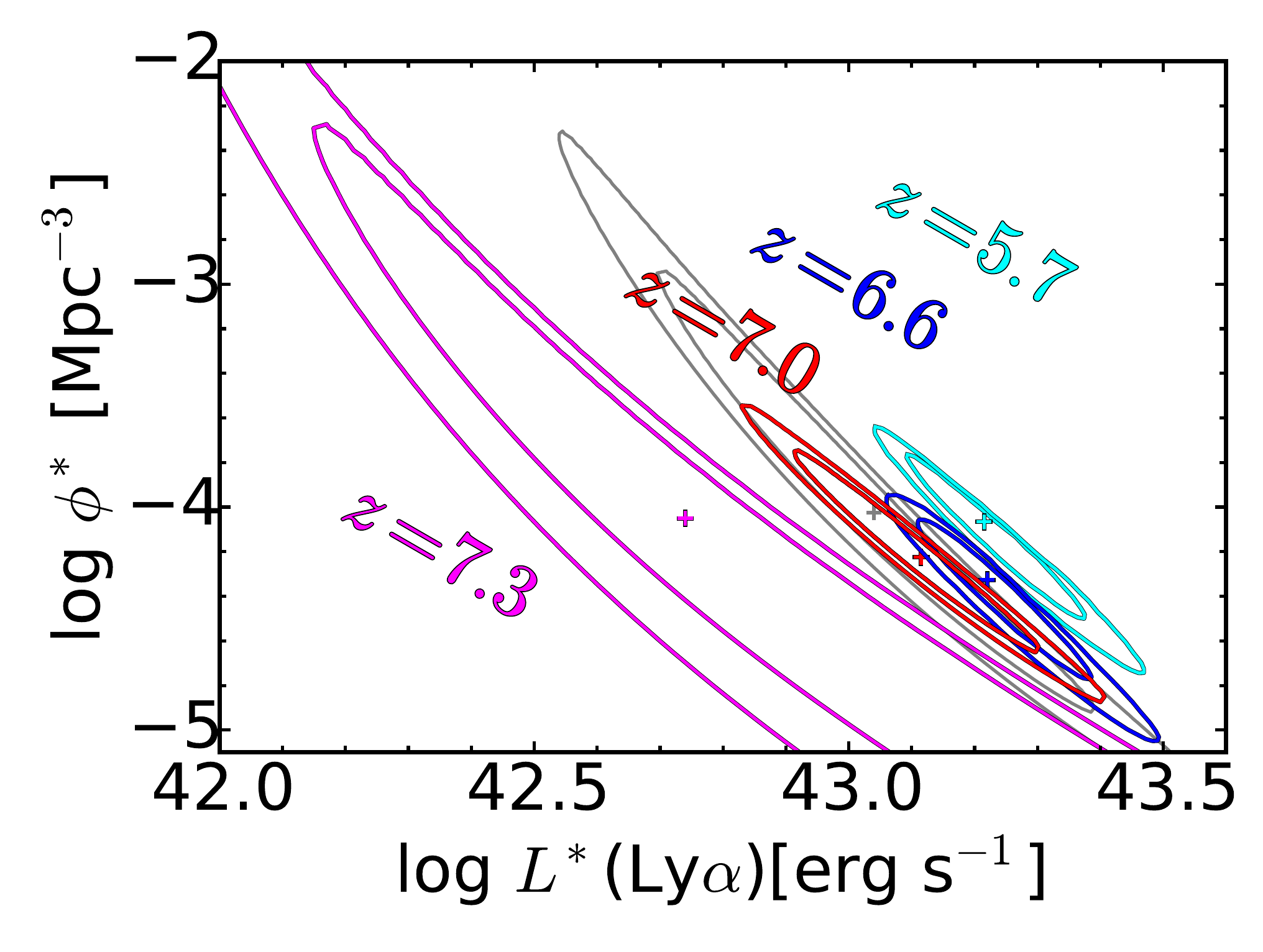}
\caption{
Same as Figure \ref{fig:contour1} but for the Ly$\alpha$ LFs at $z=5.7-7.3$. 
The red (gray) contours represent the fit to our $z=7.0$ LF 
with (without) the two data points in \cite{ota2017}
for the fixed slope of $\alpha=-2.5$. 
The cyan, blue and magenta contours denote those at $z=5.7$ \citep{konno2018}, $z=6.6$ \citep{konno2018}, 
and $z=7.3$ \citep{konno2014}, respectively.
\label{fig:contour2}}
\end{figure}

To evaluate the evolution of Ly$\alpha$ LF from $z=5.7$ to 7.3 more quantitatively, 
we investigate the error distribution of Schechter parameters.
Figure \ref{fig:contour2} presents the error contours of the Schechter parameters of Ly$\alpha$ LFs at $z=5.7$, 6.6, 7.0, and 7.3.
We fix the faint-end slopes of LFs at $z=7.0$ and $7.3$ to $\alpha=-2.5$.
The $z=7.0$ Ly$\alpha$ LF is different from those at $z=5.7$ and $7.3$ at the $>$90\% confidence levels. 
On the other hand, the error contours at $z=7.0$ overlap with those of $z=6.6$ at the $68\%$ confidence level. 
These results suggest that the Ly$\alpha$ LF evolve moderately from $z=6.6$ to $z=7.0$, 
and decrease rapidly from $z=7.0$ to $7.3$.

To quantify the decrease of the Ly$\alpha$ LF, 
we evaluate the decrease rates of $L^*$ and $\phi^*$ in a given time interval. 
The results are summarized in Table \ref{tab:pureev}.
We first investigate the case of the pure luminosity evolution. 
We conduct the fitting similar to those presented in \cite{ouchi2010} and \cite{kashikawa2011}. 
We fix $\phi^*$ of the $z=6.6$ Ly$\alpha$ LF, 
and carry out the Schechter function fitting to our $z=7.0$ Ly$\alpha$ LF. 
In this way, we estimate 
the ratio of the best-fit 
$L^*$ at $z=7.0$ to the one at $z=6.6$ ($L^*_{z=7.0}/L^*_{z=6.6}$). 
We then obtain the decrease rate of 
$L^*$ from $z=6.6$ to $7.0$ that is defined by 
$\Delta L^*/\Delta t=(1-L^*_{z=7.0}/L^*_{z=6.6})/\Delta t$. 
We also fit the Schechter function to the $z=7.3$ Ly$\alpha$ LF 
with the fixed $L^*$ of the $z=7.0$ Ly$\alpha$ LF 
to obtain the ratio of the best-fit $L^*$ and the decrease rate of $L^*$ at $z=7.0-7.3$.  
We show the results for the pure luminosity evolution case
in columns 3 and 4 of Table \ref{tab:pureev}. 
The decrease rates of $L^*$ at $z=6.6-7.0$ and $z=7.0-7.3$ are 
$\Delta L^*/\Delta t=1.67$ and $13.8$, respectively. 

We obtain the results for the case of the pure number evolution 
with a similar procedure.
We perform fitting of $\phi^*$ to the $z=7.0\ (7.3)$ Ly$\alpha$ LF, 
fixing $L^*$ to the one of the $z=6.6\ (7.0)$ Ly$\alpha$ LF. 
In the number evolution case, 
the decrease rate of $\phi^*$ 
at $z=z_1-z_2$ is defined by 
$\Delta \phi^*/\Delta t=(1-\phi^*_{z_2}/\phi^*_{z_1})/\Delta t$.
We show the results for the pure number evolution case in columns 5 and 6 of Table \ref{tab:pureev}. 
The decrease rates of $\phi^*$ at $z=6.6-7.0$ and $z=7.0-7.3$ are 
$\Delta \phi^*/\Delta t=3.33$ and $21.7$, respectively. 

In both the pure luminosity and pure number evolution cases, 
the decrease rates of $L^*$ and $\phi^*$ at a given time interval increase towards higher redshift.
This suggests the increase of the neutral hydrogen fraction
towards higher redshift.

\begin{deluxetable}{cccccc}
\tablecolumns{4}
\tabletypesize{\scriptsize}
\tablecaption{Pure luminosity and number evolutions of the Schechter parameters at $z=6.6-7.3$
\label{tab:pureev}}
\setlength{\tabcolsep}{0.0in}
\tablewidth{0pt}
\tablehead{
\colhead{Redshift} & 
\colhead{$\Delta t$\tablenotemark{a}} &
\colhead{$L^*_{z_2}/L^*_{z_1}$\tablenotemark{b}} &
\colhead{$\Delta L^*/\Delta t$\tablenotemark{c}} & 
\colhead{$\phi^*_{z_2}/\phi^*_{z_1}$\tablenotemark{d}} &
\colhead{$\Delta \phi^* / \Delta t$\tablenotemark{e}} \\
\colhead{$z_1 - z_2$} & 
\colhead{(Myr)} &
\colhead{} &
\colhead{(Gyr$^{-1}$)} & 
\colhead{} &
\colhead{(Gyr$^{-1}$)} \\
}
\startdata
6.6-7.0 & 60 & 0.90 & 1.67 & 0.80 & 3.33\\ 
7.0-7.3 & 40 & 0.45 & 13.8 & 0.14 & 21.7
\enddata
\tablenotetext{a}{
Cosmic time interval in Myr corresponding to the redshift interval of $z_1-z_2$}
\tablenotetext{b}{
Ratio of the best-fit $L^*$ at $z=z_2$ to the one at $z=z_1$ in the case of the pure luminosity evolution.}
\tablenotetext{c}{
Rate of the decrease of the best-fit $L^*$ 
in the redshift interval of $z_1-z_2$ defined as 
$\Delta L^*/\Delta t=(1-L^*_{z_2}/L^*_{z_1})/\Delta t$.
}
\tablenotetext{d}{
Ratio of the best-fit $\phi^*$ at $z=z_2$ to the one at $z=z_1$ in the case of the pure luminosity evolution.}
\tablenotetext{e}{
Rate of the decrease of the best-fit $\phi^*$ 
in the redshift interval of $z_1-z_2$ defined as 
$\Delta \phi^*/\Delta t=(1-\phi^*_{z_2}/\phi^*_{z_1})/\Delta t$.
}
\end{deluxetable}

\subsection{Evolution of Ly$\alpha$ Luminosity Densities and Cosmic Reionization}
\label{sec:evolution_of_lyald}
In this section, we discuss the implications for the cosmic reionization 
based on our Ly$\alpha$ LF. 
We derive the two quantities of the Ly$\alpha$ luminosity density (LD) and the Ly$\alpha$ transmission fraction. 
Then we compare the two quantities with the reionization models to 
estimate the neutral hydrogen fraction, $x_{\rm HI}$, at $z=7.0$.
The procedure of estimating the neutral hydrogen fraction is 
similar to those of the previous Ly$\alpha$ LF studies \citep{ouchi2010, konno2014, konno2018, ota2017, zhengy2017}. 

\begin{figure}
\epsscale{1.25}
\plotone{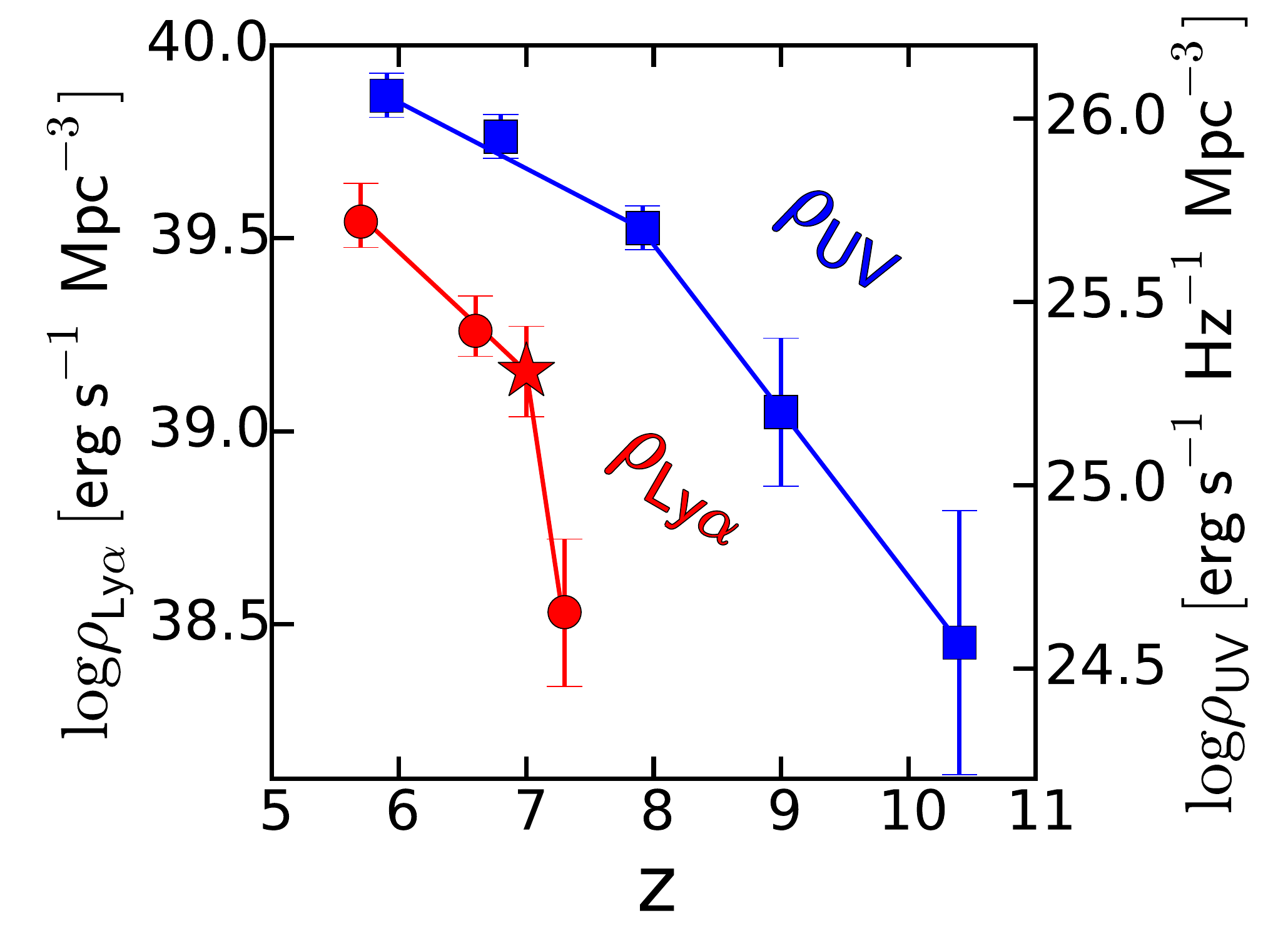}
\caption{
Redshift evolution of the Ly$\alpha$ and UV LDs obtained with LAE and LBG samples.
The red filled star-mark represents the Ly$\alpha$ LDs at $z=7.0$ from this study in the case of $\alpha=-2.5$.
The red filled circles indicate  the Ly$\alpha$ LDs at $z=5.7$, 6.6, and 7.3 \citep{konno2014, konno2018}.
The blue squares are the UV LDs given by \cite{bouwens2015b} for $z=5.9,\ 6.8,\ 7.9,\ 9.0,$ and 10.4, and \cite{ellis2013} for $z=9.0$ 
(see also Figure 11 of \citealt{konno2014}).
In this figure, we adopt the integration limits of 
$\log L_{{\rm Ly\alpha}}[{\rm erg\ s^{-1}}]=42.4$ and 
$M_{\rm UV}=-17$ for the Ly$\alpha$ and UV LD estimates.
\label{fig:LD}}
\end{figure}

We calculate the Ly$\alpha$ luminosity densities (LDs), $\rho^{{\rm Ly\alpha}}$, 
down to the luminosity of $\log L_{{\rm Ly\alpha}}[{\rm erg\ s^{-1}}]=42.4$
that corresponds to the flux limit 
for the previous surveys of LAEs at $z=5.7-7.3$. 
Figure \ref{fig:LD} represents the evolution of the $\rho^{{\rm Ly\alpha}}$.
We obtain the error bars of the Ly$\alpha$ LD, 
calculating the maximum and minimum values of the Ly$\alpha$ LD 
using the Schechter parameters $L^*$ and $\phi^*$ in the $1\sigma$ error range. 

We compare the Ly$\alpha$ LDs at $z=7.0$ (in the EoR) and $z=5.7$ (at the post reionization epoch) to estimate $T^{{\rm IGM}}_{{\rm Ly\alpha}, z=7.0}/T^{{\rm IGM}}_{{\rm Ly\alpha}, z=5.7}$, where $T^{{\rm IGM}}_{{\rm Ly\alpha}, z}$ is Ly$\alpha$ transmission through the IGM at the redshift $z$. 
The observed Ly$\alpha$ LD ($\rho^{{\rm Ly\alpha}}$) can be obtained from

\begin{eqnarray}
\rho^{{\rm Ly\alpha}} = \kappa \  T^{{\rm IGM}}_{{\rm Ly\alpha}, z} \ f^{{\rm esc}}_{{\rm Ly\alpha}} \ \rho^{{\rm UV}}.
\label{eq:rholya}
\end{eqnarray}
$\kappa$ is a conversion factor from UV to Ly$\alpha$ luminosities. 
$f^{{\rm esc}}_{\rm Ly\alpha}$ is the Ly$\alpha$ escape fraction through the interstellar medium (ISM) of a galaxy. 
$\rho^{{\rm UV}}$ is the intrinsic UV LD. 
Assuming that $\kappa$ and $f^{{\rm esc}}_{\rm Ly\alpha}$ do not evolve from $z=5.7$ to $7.0$, we obtain

\begin{eqnarray}
\frac{T^{{\rm IGM}}_{{\rm Ly\alpha}, z=7.0}}{T^{{\rm IGM}}_{{\rm Ly\alpha}, z=5.7}} = \frac{\rho^{{\rm Ly\alpha}}_{z=7.0}/\rho^{{\rm Ly\alpha}}_{z=5.7}}{\rho^{{\rm UV}}_{z=7.0}/\rho^{{\rm UV}}_{z=5.7}}.
\label{eq:rholya}
\end{eqnarray}

We estimate the ratio of UV LDs with dropouts, 
assuming that the Ly$\alpha$ emission of LAEs is originated 
from the star formation.
In this assumption, 
LAEs are the subsample of dropouts. 
We apply the ratio of UV LDs $\rho^{{\rm UV}}_{z=7.0}/\rho^{{\rm UV}}_{z=5.7}=0.57 \pm 0.07$ 
obtained by the UV LFs of \citep{bouwens2015b}. 
Here, we integrate the UV LFs down to $M_{{\rm UV}}=-17$ mag, the observed magnitude limit of \citep{bouwens2015b}, to estimate the UV LDs. 
Based on the $z=5.7$ Ly$\alpha$ LF taken from \cite{konno2018}, 
we estimate the Ly$\alpha$ LD at $z=5.7$ to be $\rho^{{\rm Ly\alpha}}_{z=5.7}=3.49 \times 10^{39}\ {\rm erg\ s^{-1}\ Mpc^{-3}}$
(Table \ref{tab:schechter_param}). 
We thus obtain the ratio of Ly$\alpha$ LD 
$\rho^{{\rm Ly\alpha}}_{z=7.0}/\rho^{{\rm Ly\alpha}}_{z=5.7}=0.41^{+0.15}_{-0.12}$. 
Combining the ratios of the UV and Ly$\alpha$ LDs, 
we obtain $T^{{\rm IGM}}_{{\rm Ly\alpha}, z=7.0}/T^{{\rm IGM}}_{{\rm Ly\alpha}, z=5.7} = 0.72^{+0.28}_{-0.22}$
with the Equation \ref{eq:rholya}.

We use theoretical models to constrain $x_{\rm HI}$ at $z=7.0$ 
with the Ly$\alpha$ LDs and the Ly$\alpha$ transmission fraction estimated above. 
We refer to theoretical models as many as possible 
to avoid the systematic uncertainties 
between different models, and to make a conservative constraint on $x_{\rm HI}$. 
We first use the analytic model of \cite{santos2004} 
to estimate the neutral hydrogen fraction $x_{{\rm HI}}$.
\cite{santos2004} assumes the galactic outflow 
with the Ly$\alpha$ velocity shifts of 0 and $360\ {\rm km\ s^{-1}}$ from the systemic velocity.
Recent studies have reported that the Ly$\alpha$ emission line at $z=2.2$ is redshifted by $\sim 200\ {\rm km\ s^{-1}}$ \citep{hashimoto2013, shibuya2014}.
Based on Figure 25 of \cite{santos2004}, 
we find that our Ly$\alpha$ transmission fraction estimate is consistent 
with the model of $x_{{\rm HI}}\leq0.5$, 
including the two velocity shift cases.

Next, we apply the combination of two theoretical models 
to estimate $x_{{\rm HI}}$. 
\cite{dijkstra2007} calculate 
the Ly$\alpha$ transmission fraction 
$T^{{\rm IGM}}_{{\rm Ly\alpha}, z=6.5}/T^{{\rm IGM}}_{{\rm Ly\alpha}, z=5.7}$ 
as a function of typical radius of ionized bubbles at $z=6.5$ 
with two cases where the ionizing background is 
or is not boosted by undetected sources around LAEs. 
Under the assumption that the characteristic size of ionized bubbles 
does not evolve between $z=6.5$ and 7.0 at a fixed $x_{{\rm HI}}$,
their model suggests a typical ionized bubble size of $\geq 8$ comoving Mpc 
for our result of 
$T^{{\rm IGM}}_{{\rm Ly\alpha}, z=7.0}/T^{{\rm IGM}}_{{\rm Ly\alpha}, z=5.7}= 0.72^{+0.28}_{-0.22}$. 
Using the relation between the typical bubble radius and $x_{{\rm HI}}$  
derived by the \cite{furlanetto2006} model (see the long dashed line in their Figure 1), 
we estimate the neutral hydrogen fraction to be $x_{{\rm HI}}\leq0.3$ at $z=7.0$. 

We also compare our Ly$\alpha$ LF with the prediction from radiative transfer simulations of \cite{mcquinn2007}.
\cite{mcquinn2007} calculate the cumulative Ly$\alpha$ LF with different values of $x_{{\rm HI}}$.
Based on Figure 4 of \cite{mcquinn2007}, we obtain $x_{{\rm HI}}= 0-0.4$ at $z=7.0$.

Finally, we use the cosmological simulation of \cite{inoue2018} 
who presented the first LAE model simultaneously reproducing all 
observational data at $z\sim6$, 
namely LAE LFs, 
LAE angular correlation functions, 
and LAE fractions in LBGs at $z>6$. 
\cite{inoue2018} derive the relation between $x_{{\rm HI}}$ and 
a ratio of the observed to the intrinsic Ly$\alpha$ LDs. 
Referring to Figure 19 of \cite{inoue2018}, we obtain $x_{{\rm HI}}\leq0.4$ 
with our result of the Ly$\alpha$ LD, 
$\rho^{{\rm Ly\alpha}}_{z=7.0}=1.43^{+0.45}_{-0.33}\times10^{39}\ {\rm erg\ s^{-1}\ Mpc^{-3}}$ 
(Table \ref{tab:schechter_param}).

In the discussion above, 
we adopt the integration limits of 
$\log L_{{\rm Ly\alpha}}[{\rm erg\ s^{-1}}]=42.4$ and 
$M_{\rm UV}=-17$ for the Ly$\alpha$ and UV LD estimates.
We check the systematic uncertainties 
raised by the choice of the integration limits.
If we change the integration limit of the Ly$\alpha$ LDs 
over $\log L_{{\rm Ly\alpha}}[{\rm erg\ s^{-1}}]=42.4-41.0$, 
the ratio of the Ly$\alpha$ LDs 
$\rho^{{\rm Ly\alpha}}_{z=7.0}/\rho^{{\rm Ly\alpha}}_{z=5.7}$ 
falls in the range of $0.40-0.41$, 
indicative of no significant difference. 
We check the values of the UV LDs
with the integration limit of 
$M_{\rm UV}=-15$, the observed limit 
in Hubble Frontier Fields \citep{atek2015,ishigaki2018}. 
Based on this integration limit, 
the ratio of $\rho^{{\rm UV}}_{z=7.0}/\rho^{{\rm UV}}_{z=5.7}=0.64 \pm 0.07$ is obtained.
We find the value of 
$T^{{\rm IGM}}_{{\rm Ly\alpha}, z=7.0}/T^{{\rm IGM}}_{{\rm Ly\alpha}, z=5.7} 
= 0.64^{+0.24}_{-0.18}$. 
The neutral hydrogen fraction is estimated to be $x_{\rm HI}=0-0.55$ 
from the model of \cite{santos2004}, 
and $x_{\rm HI}=0-0.4$ from the combination of the models of 
\cite{dijkstra2007} and \cite{furlanetto2006}. 
These values of $x_{\rm HI}$ are comparable with those obtained by the
integration limit down to $M_{\rm UV}=-17$ ($x_{\rm HI}=0-0.5$).
Note that \cite{mcquinn2007} and \cite{inoue2018} models give the same results, 
because these models do not require 
$T^{{\rm IGM}}_{{\rm Ly\alpha}, z=7.0}/T^{{\rm IGM}}_{{\rm Ly\alpha}, z=5.7}$ 
to constrain the value of $x_{\rm HI}$.
We thus conclude that the choice of the integration limits 
do not change our conclusions of $x_{\rm HI}$ estimates.

\begin{figure*}
\begin{minipage}{0.5\hsize}
\begin{center}
\epsscale{1.2}
\plotone{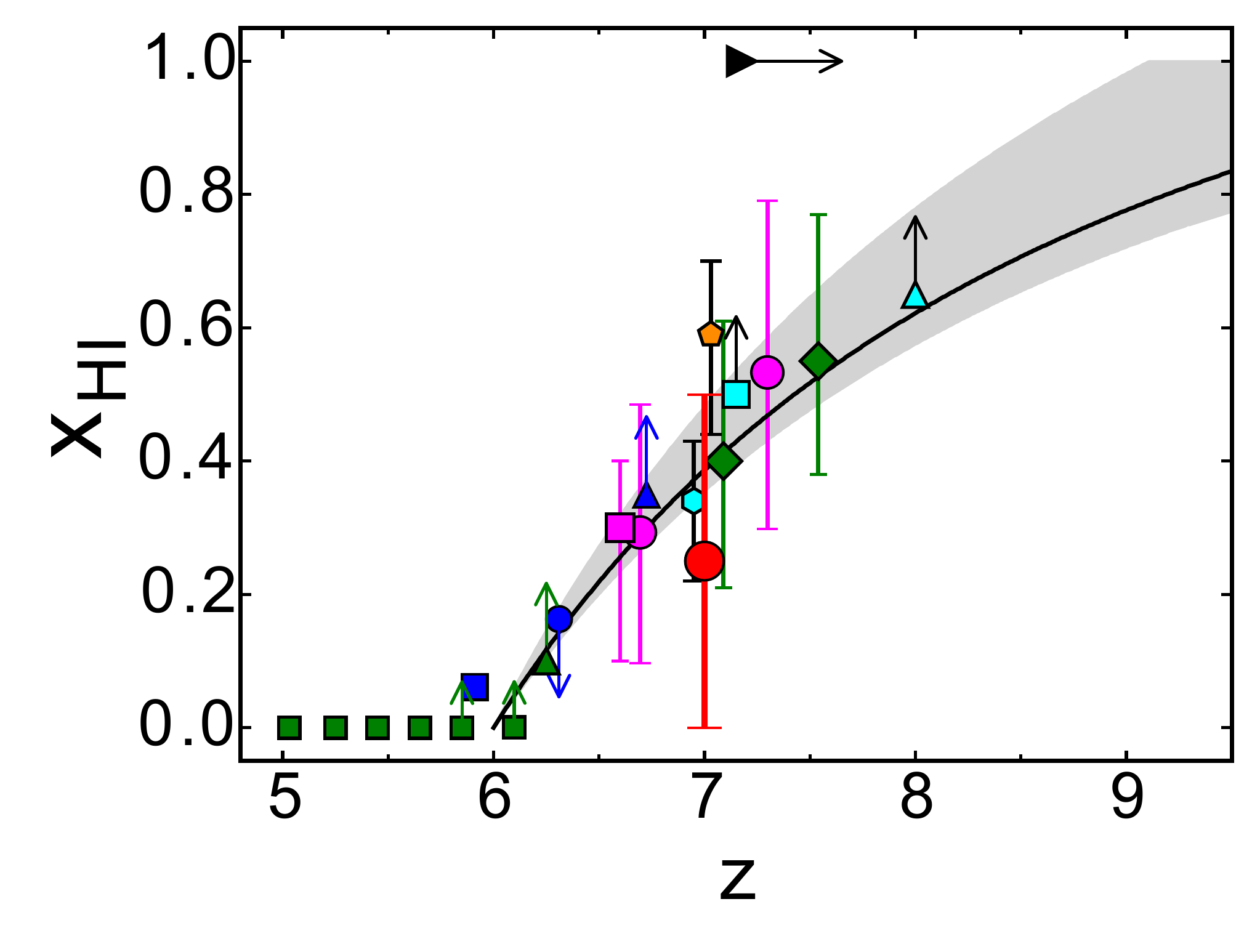}
\end{center}
\end{minipage}
\begin{minipage}{0.5\hsize}
\begin{center}
\epsscale{1.2}
\plotone{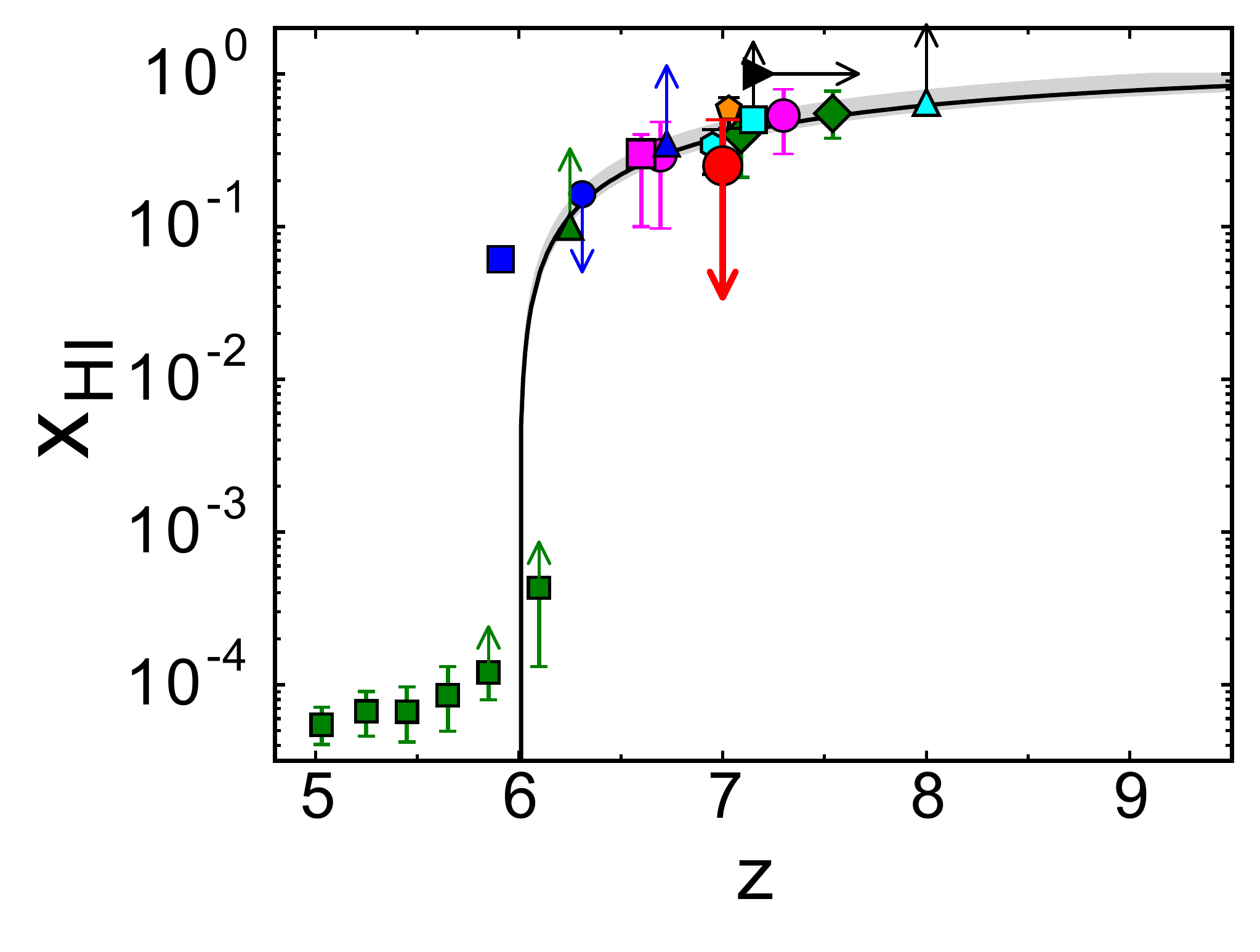}
\end{center}
\end{minipage}
\caption{
Left: Neutral hydrogen fraction $x_{{\rm HI}}$ of the IGM as a function of redshift.
The red filled circle is the $x_{{\rm HI}}$ value at $z=7.0$ estimated by our study. 
The magenta filled circles are the $x_{{\rm HI}}$ values 
at $z=6.6$ and $7.3$ estimated from the Ly$\alpha$ LF evolution 
of \cite{konno2018} and \cite{konno2014}, respectively.
The magenta filled square indicates the $x_{{\rm HI}}$ constraint 
given by the HSC LAE clustering analysis of \cite{ouchi2018}.
The blue triangle, circle, and square denote the results from the Ly$\alpha$ damping wing absorption of GRBs 
at $z=5.9$ \citep{totani2016}, $z=6.3$ \citep{totani2006}, and $z=6.7$ \citep{greiner2009}, respectively.
The green filled squares are the constraints from the QSO Gunn-Peterson optical depth measurement results \citep{fan2006}.
The green filled diamonds represent the $x_{{\rm HI}}$ value obtained by 
damping wing absorption measurements of QSOs at $z=7.1$ and $7.5$ \citep{greig2017, banados2017}. 
The result for QSO at $z=6.3$ \citep{schroeder2013} is also shown with the green filled triangle. 
The orange pentagon shows the $x_{{\rm HI}}$ estimate at $z\sim7$ provided by \cite{mason2017} 
based on the model to infer $x_{{\rm HI}}$ from 
the observed EW distribution of Ly$\alpha$ emission from LBGs.
The cyan square indicates the $x_{{\rm HI}}$ constraint 
from the fraction of Ly$\alpha$ emitting LBGs at $z\sim7$ 
(the combined constraint from \citealt{stark2010, pentericci2011, pentericci2014, caruana2012, caruana2014, ono2012, schenker2012, furusawa2016}), 
while the cyan hexagon and triangle
are the results from \cite{schenker2014} at $z\sim7$ and $8$.
The black triangle shows the $1\sigma$ lower limit of the redshift obtained by
\cite{planck2016b} in the case of instantaneous reionization. 
The solid line and the gray shade indicate the $x_{{\rm HI}}$ evolution and uncertainties 
estimated from $\rho_{{\rm UV}}$ analysis \citep{ishigaki2018}. 
Right: Same as the left panel, but for the log scale ordinate axis.
\label{fig:xHI}}
\end{figure*}

Based on the results described above, 
we conclude that the neutral hydrogen fraction is estimated to be $x_{{\rm HI}}\leq0.5$, i.e. $x_{{\rm HI}}=0.25\pm0.25$ at $z=7.0$,  
taking the most conservative value. 
In our neutral hydrogen fraction estimate, 
we include the variance of the theoretical models 
and the uncertainties in our Ly$\alpha$ transmission fraction estimates. 
Figure \ref{fig:xHI} shows our estimate of $x_{\rm HI}$ at $z=7.0$ and those taken from previous studies. 
The $x_{{\rm HI}}$ measurement of our result is consistent with those derived by 
the QSO damping wing study \citep{greig2017}, 
the Ly$\alpha$ EW analysis \citep{mason2017}, 
and the $\rho_{{\rm UV}}$ evolution work \citep{ishigaki2018} 
within uncertainties. 

\section{Conclusions}
\label{sec:conclusions}
We conduct an ultra-deep and large-area HSC imaging survey 
with the $NB973_{\rm HSC}$ filter under the CHORUS project.
We observe a total of $3.1\ {\rm deg^2}$ area sky consisting of two independent blank fields, COSMOS and SXDS. 
We have identified 34 LAE candidates at $z=7.0$, 
and made the largest sample of $z=7.0$ LAEs, to date. 
Our survey volume is large enough to investigate 
the existence of the bright-end excess of the Ly$\alpha$ LF.
The major results of our study are summarized below.
\begin{enumerate}
\item
Based on our LAE sample, 
we derive the Ly$\alpha$ LF at $z=7.0$ at the luminosity range of $\log L_{{\rm Ly\alpha}}[{\rm erg\ s^{-1}}]=42.9-43.6$.
We compare our Ly$\alpha$ LF 
with the previous measurements of Ly$\alpha$ LFs at $z=7$. 
Our number densities 
are consistent with that of \cite{zhengy2017} and \cite{ota2017} 
at the bright end ($\log L_{{\rm Ly\alpha}}[{\rm erg\ s^{-1}}]=43.3-43.6$) 
and faint end ($\log L_{{\rm Ly\alpha}}[{\rm erg\ s^{-1}}]=42.9-43.3$), respectively. 
We find that the shape of the $z=7.0$ Ly$\alpha$ LF can be explained by the Schechter function, 
and that there is no clear signature of a bright-end excess 
over the best-fit Schechter function at $z=7$.
\item
We compare the Ly$\alpha$ LF at $z=7.0$ with those at $z=5.7$, $6.6$, and $7.3$.
Our Ly$\alpha$ LF show a weak decrease from the one at $z=6.6$. 
The Ly$\alpha$ LF at $z=7.0$ shows a clear decrease from the one at $z=7.3$. 
We find that $L^*$ and $\phi^*$ decrease acceleratingly toward high redshifts
in both pure luminosity and number evolution cases.
\item
Comparing the redshift evolutions of Ly$\alpha$ LD and UV LD, 
we estimate the IGM transmission of Ly$\alpha$ photons to be 
$T^{{\rm IGM}}_{{\rm Ly\alpha}, z=7.0}/T^{{\rm IGM}}_{{\rm Ly\alpha}, z=5.7} = 0.72^{+0.28}_{-0.22}$
with the Ly$\alpha$ LDs and the UV LDs estimated with dropouts. 
We compare the IGM transmission estimate with several different reionization models, 
and obtain the neutral hydrogen fraction estimate $x_{{\rm HI}}=0.25\pm0.25$ at $z=7.0$.
\end{enumerate}

\acknowledgments
We are grateful to 
Richard Ellis,  
Hisanori Furusawa, 
Ryo Higuchi, 
Weida Hu, 
Shotaro Kikuchihara, 
Takashi Kojima, 
Hilmi Miftahul, 
Shiro Mukae, 
Yukie Oishi, 
Yuma Sugahara, 
Jun Toshikawa, 
and Zheng Zheng 
for useful comments and discussions.

The Hyper Suprime-Cam (HSC) Collaboration includes the astronomical communities of Japan and Taiwan, and Princeton University.
The HSC instrumentation and software were developed by the National Astronomical Observatory of Japan (NAOJ), the Kavli Institute for the Physics and Mathematics of the Universe (Kavli IPMU), the University of Tokyo, the High Energy Accelerator Research Organization (KEK), the Academia Sinica Institute for Astronomy and Astrophysics in Taiwan (ASIAA), and Princeton University. 
Funding was contributed by the FIRST program from Japanese Cabinet Office, the Ministry of Education, Culture, Sports, Science and Technology (MEXT), the Japan Society for the Promotion of Science (JSPS), Japan Science and Technology Agency (JST), the Toray Science Foundation, NAOJ, Kavli IPMU, KEK, ASIAA, and Princeton University. 

The Pan-STARRS1 Surveys (PS1) have been made possible through contributions of the Institute for Astronomy, the University of Hawaii, the Pan-STARRS Project Office, the Max Planck Institute for Extraterrestrial Physics, Garching, The Johns Hopkins University, Durham University, the University of Edinburgh, Queen’s University Belfast, the Harvard-Smithsonian Center for Astrophysics, the Las Cumbres Observatory Global Telescope Network Incorporated, the National Central University of Taiwan, the Space Telescope Science Institute, the National Aeronautics and Space Administration under Grant No. NNX08AR22G issued through the Planetary Science Division of the NASA Science Mission Directorate, the National Science Foundation under Grant No. AST-1238877, the University of Maryland, and Eotvos Lorand University (ELTE). 

This paper makes use of software developed for the Large Synoptic Survey Telescope. We thank the LSST Project for making their code available as free software at http://dm.lsst.org.

Based in part on data collected at the Subaru Telescope and retrieved from the HSC data archive system, which is operated by Subaru Telescope and Astronomy Data Center at National Astronomical Observatory of Japan.

This research has benefitted from the SpeX Prism Spectral Libraries, maintained by Adam Burgasser at http://pono.ucsd.edu/~adam/browndwarfs/spexprism.

The NB718 and NB816 filters were supported by Ehime University (PI: Y. Taniguchi). 
The NB921 and NB973 filters were supported by KAKENHI (23244025) Grant-in-Aid for Scientific Research (A) through the Japan Society for the Promotion of Science (PI: M. Ouchi). 
This work is supported by World Premier International Research Center
Initiative (WPI Initiative), MEXT, Japan, and KAKENHI (15H02064)
Grant-in-Aid for Scientific Research (A) through Japan Society for the
Promotion of Science.
R.I. acknowledges support from the Advanced Leading Graduate Course for Photon Science (ALPS) grant. 
Y.T. is supported by JSPS KAKENHI Grant Number 16H02166. 

\bibliographystyle{apj}
\bibliography{apj-jour,itoh2018}

\end{document}